\pdfoutput=1
\documentclass{IEEEtran}
\usepackage{cite}
\usepackage{amsmath,amssymb,amsfonts}
\usepackage{graphicx}
\usepackage{textcomp}
\def\BibTeX{{\rm B\kern-.05em{\sc i\kern-.025em b}\kern-.08em
    T\kern-.1667em\lower.7ex\hbox{E}\kern-.125emX}}
    
\usepackage{hyperref}
\usepackage{braket}
\usepackage{subcaption}
\usepackage{tikz}
\usepackage{breqn}
\usepackage{algorithm}
\usepackage[noend]{algpseudocode}
\usepackage{qcircuit}    
\usepackage{comment}   
\usepackage{float}

\begin{document}

\newcommand{\ahsan}[1]{[{\color{red}Ahsan: #1}]}
\newcommand{\sumsam}[1]{[{\color{cyan}Sumsam: #1}]}

\title{K-Means Clustering on Noisy Intermediate Scale Quantum Computers}
\author{
  \IEEEauthorblockN{Sumsam Ullah Khan$^{1,2}$, Ahsan Javed Awan$^1$ and Gemma Vall-Llosera$^1$}\\
  \IEEEauthorblockA{$^1$Department of Cloud Systems and Platforms, Ericsson Research, Sweden\\
  $^2$School of Electrical Engineering and Computer Science, Royal Institute of Technology, Sweden\\
   \{sumsam.khan, ahsan.javed.awan, gemma.vall-llosera\}@ericsson.com}
}
\maketitle

\begin{abstract}
Real-time clustering of big performance data generated by the telecommunication networks requires domain-specific high performance compute infrastructure to detect anomalies. In this paper, we evaluate noisy intermediate-scale quantum (NISQ) computers characterized by low decoherence times, for K-means clustering and propose three strategies to generate shorter-depth quantum circuits needed to overcome the limitation of NISQ computers. The strategies are based on exploiting; i) quantum interference, ii) negative rotations and iii) destructive interference. By comparing our implementations on IBMQX2 machine for representative data sets, we show that NISQ computers can solve the K-means clustering problem with the same level of accuracy as that of classical computers. 
\end{abstract}


%
\IEEEpeerreviewmaketitle

\section{Introduction}
\label{sec:introduction}

Machine Learning is being widely adopted by the industry to transform the massive amount of data into insights that drive the growth of the companies in the era of digital transformation, e.g. supervised and unsupervised machine learning algorithms like SVM, K-Means, Linear regression, etc. are being used for automation of network management offering. The combination of compute intensity of machine learning algorithms and the massive data volumes has spurred renewed interest in the specialized high-performance computing platforms. 

Quantum computers are one of them that exploit the principles of quantum mechanics: i) Quantisation: energy, momentum, angular momentum and other physical quantities of a bound system are restricted to discrete values (quantised); ii) Wave-particle duality: objects are both waves and particles; iii) Heisenberg principle: the more precisely the position of some particle is determined, the less precisely its momentum can be known, and vice versa; thus there is a fundamental limit to the measurement precision of physical quantities of a particle; iv) Superposition: two quantum states can be added together and the result is another valid quantum state; v) Entanglement: when the quantum state of any particle belonging to a system cannot be described independently of the state of the other particles, even when separated by a large distance, the particles are entangled; vi) Fragility: by measuring a quantum system we destroy any previous information. From this, it follows the no-cloning theorem that states: it is impossible to create an identical copy of an arbitrary unknown quantum state. Executing machine learning algorithms on a quantum computer requires to re-design the classical algorithms so they are bound to the principles of quantum mechanics.

Current and upcoming quantum computers also termed \textit{Noisy
Intermediate Scale Quantum computers (NISQ)} do not provide sufficient fault tolerance~\cite{preskill2018quantum}. Qubits on the current quantum devices have low coherence times. This leads to decoherence and errors in the computation \cite{qubits:2018}. Algorithm implementations with shallow depth quantum circuits can provide us with better results on NISQ computers considering that the complexity of implementation is measured as the total number of elementary gates required to build the circuit \cite{Adriano:1995}. In this paper, we explore the quantum implementation of K-means clustering algorithm and propose three optimization strategies to achieve shorter-depth circuits for quantum K-means on NISQ computers. In particular, we make the following contributions:

\begin{itemize}
\item We provide an implementation of K-means clustering algorithm using a shallow depth quantum interference circuit which prepares the quantum states of qubit based on the input vectors such that the angle between the interfering copies of the input vectors is equal to the angle between the interfering copies of the vectors in the final quantum state. 
\item We propose a method referred to as \textit{Negative Rotations}, which maps the cosine similarity of two vectors on the probability of the $\ket{0}$ state of the qubit. The implementation of K-means algorithm which uses the probability $P\ket{0}$ as a metric to assign clusters to test vectors is provided. 
\item We propose a method for calculating distances using the destructive interference probabilities of the quantum state which is prepared based on the input vectors. An implementation of the circuit that calculates distances between two vectors used to realize the K-means algorithm is also given.
\end{itemize}

This paper is structured as follows. First we summarise the basic concepts of quantum computing and the K-means algorithm (section \ref{sec:background}). Then we discuss the related work in the field (section \ref{sec:relatedwork}). We continue with our implementations for quantum K-means (section \ref{sec:qkmeans_implementations}), how we have evaluated our implementations (section \ref{sec:evaluation}). It follows the results we obtained with a thorough discussion (section \ref{sec:results}) and we finalise with the conclusions (section \ref{sec:conclusions}).




\section{Background}  \label{sec:background}

\subsection{K-Means} \label{sec:kmeans}
K-means~\cite{kmeans:1982:lloyd} is an unsupervised clustering algorithm that groups together \texttt{n} observations into \texttt{K} clusters making sure intra cluster variance is minimized and inter set variance is maximized. Training vectors are assigned to the cluster of the nearest centroid iteratively. New centroids are calculated at the end of the iterations by averaging the vectors belonging to the corresponding clusters. The time complexity of the classical version of the algorithm is dependent on the number of features in the input vectors $N$, the total number of input vectors $M$, and $K$ (number of clusters) $\mathcal{O}(M N K)$

\subsection{Noisy Intermediate Scale Quantum (NISQ) Computers}
Noisy Intermediate Scale Quantum (NISQ) Computers \cite{NISQ:2018} are the near future quantum computers with 50-100 qubits. These computers are potentially capable of solving tasks exponentially faster than today's classical computers. However, there are certain limitations that restrict the effectiveness of NISQ computers. The noise in the quantum gates and low coherence times of the qubits will limit the size of the quantum circuit that can be executed. Keeping in view these limitations, smaller depth quantum circuits are desired so they can be executed reliably on NISQ computers. 

\subsection{Basic Quantum Computing Concepts}

\subsubsection{Qubit}
Qubit is the smallest unit of information in a quantum computer. Binary bit states are written using Dirac Vector or Bra-Ket notation as follows, $\ket{0} = \bigl( \begin{smallmatrix}1\\0\end{smallmatrix}\bigr)$ and $\ket{1} = \bigl( \begin{smallmatrix}0\\1\end{smallmatrix}\bigr)$.

Unlike a classical bit which can be in either state $\ket{0}$ or in state $\ket{1}$, a qubit can be in state $\ket{0}$ and state $\ket{1}$ at the same time with some probability of being in either state. The state of a qubit is represented as, $\ket{\psi} = \bigl( \begin{smallmatrix}a\\b\end{smallmatrix}\bigr)$
where $a$ and $b$ are complex numbers and $|a|^2 + |b|^2 = 1$. These are the probability amplitudes for the qubit being in either state. $|a|^2$ is the probability of finding the qubit after measurement in state $\ket{0}$ and $|b|^2$ is the probability for state $\ket{1}$. The state of a qubit can also be written as $\ket{\psi} = a \ket{0} + b \ket{1}$.

A qubit in such a state is said to be in superposition, so when a measurement is done on that qubit, it would yield $\ket{0}$ with probability $|a|^2$ and $\ket{1}$ with probability $|b|^2$.

Multiple bits in dirac vector notation are represented as tensor products. So a bit string $01$ is written as 

\begingroup
\fontsize{8pt}{8pt}\selectfont
\[ \ket{0}  \otimes \ket{1} = \begin{pmatrix}1\\0\end{pmatrix} \otimes \begin{pmatrix}0\\1\end{pmatrix} = \begin{pmatrix}1\begin{pmatrix}0\\1\end{pmatrix}\\0\begin{pmatrix}0\\1\end{pmatrix}\end{pmatrix} =  \begin{matrix}\ket{00}\\\ket{01}\\\ket{10}\\\ket{11}\end{matrix} \begin{pmatrix}0\\1\\0\\0\end{pmatrix}   \]
\endgroup

Similarly multiple qubits are represented as tensor products of individual qubits, but instead of a definite state, the tensor product state holds the probability amplitudes of each possible state. A two qubit tensor product state is written as, 

\begingroup
\fontsize{8pt}{8pt}\selectfont
\begin{equation}
  \ket{\psi} = \begin{pmatrix}a_0\\a_1\\a_2\\a_3 \end{pmatrix} = a_0 \ket{00} + a_1 \ket{01} + a_2 \ket{10} + a_3 \ket{11}\
\label{eq:str1.1}
\end{equation}
\endgroup

The product state is normalized so that, \[|a_0|^2 + |a_1|^2 + |a_2|^2 + |a_3|^2 = 1\]

\subsubsection{Entanglement}
If two qubits are entangled it means that measuring one qubit collapses the superposition of the other qubit as well. An entangled state is a multi-qubit quantum state that can not be written as a Kronecker product of single-qubit states.
For example, the individual qubits that result on the following product state are entangled 
\[  \begin{bmatrix} \frac{1}{\sqrt{2}}&0&0&\frac{1}{\sqrt{2}} \end{bmatrix}^T \]
 as individual qubits states cannot be factored out from this product:

\begin{table}[H]
\centering
    \begin{tabular}{cc}
        $\begin{pmatrix} \frac{1}{\sqrt{2}}\\0\\0\\ \frac{1}{\sqrt{2}} \end{pmatrix} = \begin{pmatrix} a\\b \end{pmatrix} \otimes \begin{pmatrix} c\\d \end{pmatrix} \xrightarrow{}$ &
        $\begin{matrix} ac= \frac{1}{\sqrt{2}} \\ ad= 0\\bc= 0\\bd= \frac{1}{\sqrt{2}} \end{matrix}$
    \end{tabular}
\end{table}

Since there is no solution to the set of equations, this means that the individual states $\bigl( \begin{smallmatrix}a\\b\end{smallmatrix}\bigr)$ and $\bigl( \begin{smallmatrix}c\\d\end{smallmatrix}\bigr)$ cannot be factored out. Therefore, the qubits are entangled. In this entangled quantum system, if we measure the first qubit to be in state $\ket{0}$, the second qubit would automatically collapse to state $\ket{0}$ as well. Similarly if the first qubit is measured to be in state $\ket{1}$, the second qubit would immediately collapse to $\ket{1}$. This entangled system has $50\%$ probability of collapsing to state $\ket{00}$ and $50\%$ probability of collapsing to $\ket{11}$.

\subsubsection{Quantum Gates}  \label{sec:quantum_gates}

Quantum gates are the quantum operations used to manipulate and transform qubit states. After applying a quantum gate the norm of the state vector should maintain its unity, meaning the sum of the squares of probability amplitudes should be equal to one. Therefore, quantum gates are represented as unitary matrices. The conjugate transpose of a unitary matrix is its inverse, and since the inverse of all unitary matrices exists, thus all quantum operations are reversible. This holds true for every quantum gate except the measurement gate which is a non-reversible operation used at the end of the computation.

The quantum gates that are used in this work are provided in Table \ref{tab:quantum_gates}. Further reading on all quantum gates can be done in ref.~\cite{nielsen2010quantum}.

\begin{table}[h]
    \centering
    \resizebox{\columnwidth}{!}{
    \begin{tabular}{c|c|c} \hline
    \textbf{Quantum Gate} & \textbf{Matrix Representation} & \textbf{Circuit Representation} \\ \hline
    Hadamard & $\frac{1}{\sqrt{2}} \begin{pmatrix}1&1\\1&-1\end{pmatrix}$ & $\Qcircuit @C=1.5em @R=2.5em{&\gate{H} & \qw}$\\ \hline
    NOT & $\begin{pmatrix}0&1\\1&0\end{pmatrix}$ & $\Qcircuit @C=1.5em @R=2.5em{& \gate{X} & \qw}$\\ \hline
    Ry (rotation Y) & $\begin{pmatrix} \cos \frac{\theta}{2} & -\sin \frac{\theta}{2}\\\sin \frac{\theta}{2}&\cos \frac{\theta}{2}\end{pmatrix}$ & $\Qcircuit @C=1.5em @R=2.5em{& \gate{Ry(\theta)} & \qw}$\\ \hline
    CX (controlled NOT) & $\begin{pmatrix}1&0&0&0\\0&1&0&0\\0&0&0&1\\0&0&1&0\end{pmatrix}$ & $\Qcircuit @C=1.5em @R=1.5em{& \ctrl{1} & \qw \\& \targ & \qw}$ \\ \hline
    CCX (Toffoli) & $\begin{pmatrix}1&0&0&0&0&0&0&0\\0&1&0&0&0&0&0&0\\0&0&1&0&0&0&0&0\\0&0&0&1&0&0&0&0\\0&0&0&0&1&0&0&0\\0&0&0&0&0&1&0&0\\0&0&0&0&0&0&0&1\\0&0&0&0&0&0&1&0\end{pmatrix}$ & $\Qcircuit @C=1.5em @R=1.5em @!R @!C{& \ctrl{1} & \qw \\& \ctrl{1} & \qw \\& \targ & \qw}$ \\ \hline
    \end{tabular}}
    \caption{Quantum Gates}
    \label{tab:quantum_gates}
    
\end{table}

\subsubsection{Quantum Circuit Model}
\begin{figure}[h]
\centering
\makebox[1.2\linewidth]{
\input{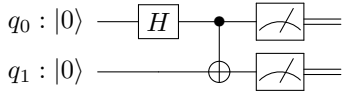}
}
\caption{Quantum Circuit for two qubit entanglement.}
\label{fig:q_circuit}
\end{figure}

Quantum computation works by performing a set of quantum operations on qubits. The quantum operations are represented by quantum gates and the quantum gates are applied to the qubits. Measurement is done at the end. This process of quantum transition of the qubits is depicted in the form of a Quantum Circuit, where the timeline of the qubit is read from left to right. Quantum Circuit model is the most famous way of developing and modelling quantum algorithms~\cite{nielsen2010quantum}. An example of a two qubit quantum circuit that prepares entangled states is given in Fig.~\ref{fig:q_circuit}. The circuit prepares this quantum state,
\[ \ket{\psi} = \frac{1}{\sqrt{2}} \ket{00} + \frac{1}{\sqrt{2}} \ket{11} \]

Fig.~\ref{fig:q_circuit} shows a quantum circuit in which a Hadamard gate is applied to the qubit $q_0$ and then a controlled-NOT gate is applied where the control is $q_0$ and the target qubit is $q_1$. At the end of the circuit a measurement gate is applied on both qubits. This circuit notation is used to illustrate quantum circuits throughout this report. 

\subsubsection{Amplitude Encoding} 
Amplitude Encoding~\cite{schuld2018supervised} is one way of encoding classical information into amplitudes of quantum states. This encoding technique is used in this work to encode classical data. Given a real value vector $v$, it is encoded into quantum information using Amplitude Encoding resulting in the following quantum state,

\begingroup
\fontsize{8pt}{8pt}\selectfont
\[ \ket{\psi} = \frac{1}{M} \sum_{i=0}^{N-1} v_i \ket{i} \ \ M = \sum_{i=0}^{N-1} \sqrt{v_i^2} \]
\endgroup

\subsubsection{Interference}
Wave-particle duality states that every quantum particle can be described as a particle as well as a wave. Waves can superpose each other producing an interference pattern. When the amplitudes of the waves are in phase we will have constructive interference (amplitudes add up) and when they are out of phase ($180^o$) we will have destructive interference (amplitudes cancel each other). 

In quantum computing, the Hadamard operation is used as an interference transformation. 
Interference of a 2 qubit system having 4 amplitudes is achieved by applying a Hadamard gate to the most significant qubit. The Hadamard gate as described in section~\ref{sec:quantum_gates} is a $2\times2$ unitary matrix. 
The matrix representation of Hadamard gate applied to the first qubit of a two qubit quantum state, can be written as a tensor product of the Hadamard matrix and the Identity matrix, since we are applying a Hadamard gate to the first qubit and not operating the second qubit.

Applying $H \otimes I$ to the state $\ket{\psi}$ results in the following,

\begingroup
\fontsize{8pt}{8pt}\selectfont
\[ \ket{\phi} = \frac{1}{\sqrt{2}} \begin{pmatrix}1&0&1&0\\0&1&0&1\\1&0&-1&0\\0&1&0&-1\end{pmatrix} \begin{pmatrix}a_0\\a_1\\a_2\\a_3 \end{pmatrix} = \frac{1}{\sqrt{2}} \begin{pmatrix}a_0+a_2\\a_1+a_3\\a_0-a_2\\a_1-a_3 \end{pmatrix} \begin{matrix}\ket{00}\\\ket{01}\\\ket{10}\\\ket{11} \end{matrix}  \]
\endgroup

The amplitude values where the first qubit is in state $\ket{0}$ interfere with the amplitude values where the first qubit is in state $\ket{1}$. This results on quantum states with constructive interference on the $\ket{0}$ state of the first qubit and destructive interference on the $\ket{1}$ state of the first qubit. We use this distinction between constructive and destructive interference to describe the solutions presented. Note that ~\cite{schuld2018supervised} describes the  Hadamard interference transformation but the authors do not exploit the destructive interference pattern.

\section{Related Work} \label{sec:relatedwork}
Here we briefly describe two approaches which are used to implement quantum K-means algorithm. Section \ref{sec:qkmeans} summarizes the approach previously published on implementing quantum K-means algorithm and in section \ref{sec:distnce_based_classifier} a recently proposed solution for distance based classification is explained.

\subsection{Quantum K-Means} \label{sec:qkmeans}
One version of the quantum K-means algorithm is described in ~\cite{QML:2018}. It uses three different quantum subroutines to perform the K-means clustering: \textit{SwapTest}, \textit{DistCalc} and \textit{Grovers Optimization}.

Overlap from the \textit{SwapTest} \cite{swapt_test:2006} subroutine is used to calculate the distance in \textit{DistCal} subroutine. The algorithm for calculating distance using \textit{SwapTest} is described in~\cite{dist_calc:2013}. The Euclidean distance $|a-b|^2$ between two vectors $a$ and $b$  is calculated by performing the following three steps. 

\begin{enumerate}
    \item \textbf{State Preparation:} Prepare two quantum states, 
    \begingroup
    \fontsize{8pt}{8pt}\selectfont
    \[\ket{\psi} = \frac{1}{\sqrt{2}} \left( \ket{0,a} + \ket{1,b} \right) \ ,\ \ket{\phi} = \frac{1}{\sqrt{Z}} \left( |a| \ket{0} + |b| \ket{1} \right)\]
    \endgroup
    where $Z = |a|^2 + |b|^2$
    \item \textbf{Find Overlap: } Calculate the overlap $\left<\psi|\phi\right>$ using \textit{SwapTest} \cite{swapt_test:2006}
    \item \textbf{Calculate Distance:} Using the following equation to calculate the Euclidean Distance
    \[ Distance = 2 Z |\left<\psi|\phi\right>|^2 \]
\end{enumerate}

Distance calculations for K-means are done using this method and the nearest centroid is found using \textit{Grovers Optimization}~\cite{grover_optimization:1996} subroutine. 

The bottleneck in the classical version of the K-means algorithm is the calculation of the distance between N dimensional vectors. Thanks to the efficient way of calculating distances using the quantum parallelism, QK-means achieves an exponential speed-up. The Lloyd's K-means algorithm~\cite{kmeans:1982:lloyd} has the time complexity $\mathcal{O}(M N K)$ while the quantum version has the time complexity $\mathcal{O}(\log(N) M K)$.

\subsection{Distance Based Classifier} \label{sec:distnce_based_classifier}
\begin{figure}[h]
    \centering
    \includegraphics[width=1\columnwidth]{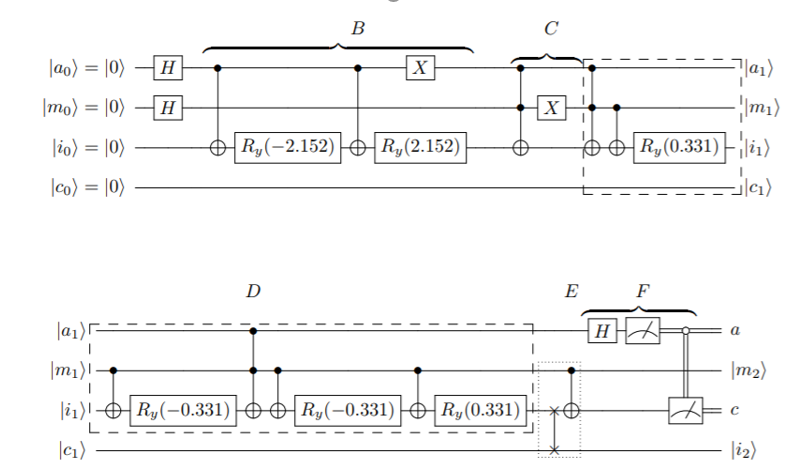}
    \caption{Distance Based Classifier Circuit for loading two training vectors and one test vector (test and training vectors taken from Iris Dataset \cite{iris:1936}).  \textbf{Step B:} Load test vector; \textbf{Step C:} Load training vector 1; \textbf{Step D:} Load training vector 2; \textbf{Step E:} Flip the class label for training vector 2; \textbf{Step F:} Interference and measurement \cite{distanceclassifier:2017:maria}. }
    \label{fig:interference}
\end{figure}

A recent study~\cite{distanceclassifier:2017:maria} proposed a quantum interference circuit that can perform distance-based classification. It uses quantum interference to compute the distance measure for the classification. 
The distance measure is calculated between  the test vector and all training vectors. This distance measure is then used for classification. Given a training dataset $ D = \{(x^1,y^1),...(x^M,y^M)\} $, with \texttt{M} training vectors having \texttt{N} dimensions $x^m \in \mathbb{R}^N$, and their corresponding class labels $y^m \in \{-1,1\}$. The goal of the classifier is to assign a class label $y^t$ to a test vector $x^t$. The binary classifier implemented by the quantum interference circuit is shown  in Fig.~\ref{fig:interference}.

The circuit in Fig.~\ref{fig:interference} implements the specific binary classification problem of the Iris dataset~\cite{iris:1936} where only two features for each sample vector are taken. The values of the angles of the rotation gates (Ry) shown in the circuit are those of the example test and training vectors presented in \cite{distanceclassifier:2017:maria}. There are two stages in the circuit: State Preparation and Interference. State Preparation loads two training vectors and one test vector (along with a copy of a test vector) on the data qubit $\ket{i_0}$ entangled with the ancilla qubit $\ket{a_0}$, index qubit $\ket{m_0}$ and class qubit $\ket{c_0}$. This state preparation continues until step E in Fig. \ref{fig:interference}. Step F constitutes a Hadamard gate applied to the ancilla qubit $\ket{a_0}$. This step performs the actual interference between test and training vectors that were loaded in the entangled state. One copy of the test vector interferes with the first training vector and the other copy interferes with the second training vector. A measurement 
operation is performed at the end of the circuit to read out the qubits' values.

This distance based classifier is only valid for datasets with two training vectors having two dimensions and it thus, cannot load arbitrary training vectors. 

\section{Our Implementations of Quantum K-Means}\label{sec:qkmeans_implementations}
Here we provide 3 different optimized implementations of quantum K-means algorithm. In section \ref{sec:constant_depth_interference} we present an implementation based on the circuit from distance based classifier (\ref{sec:distnce_based_classifier}). In section \ref{sec:negative_rotations} we propose a new solution referred to as \textit{Negative Rotations} and in section \ref{sec:destructive_interference} we provide an alternative method of calculating euclidean distances.

\subsection{Constant Depth Circuit using Interference} \label{sec:constant_depth_interference}

\subsubsection{Basic Implementation}\label{basic_implementation}
We modify the quantum circuit of distance based classifier\cite{distanceclassifier:2017:maria} shown in Fig.~\ref{fig:interference} to implement the quantum version of the K-means algorithm. One important factor that adds to the complexity of a quantum circuit is its depth. Circuit depth is the number of gates in the critical path of the circuit. The critical path is the longest  path of the circuit which includes the operations that have to be run sequentially. The metric used to measure the complexity of an implementation is the number of elementary gates required~\cite{Adriano:1995}. Reducing the number of gates required decreases the depth of the circuit and this, not only helps in achieving faster execution times but also fewer errors. 

The authors in~\cite{distanceclassifier:2017:maria}  use a specialized case for the experimental implementation of the algorithm. They use a dataset with two training vectors having 2 dimensions: $D = {(x^0,y^0),(x^1,y^1)}$ with associated values: $x^0=(0,1), y^0=-1$ and $x^1=(0.789,0.615), y^1=1$. The first training vector $x^0=(0,1)$ allows them to use a Toffoli (Controlled Controlled Not) gate to load the training vector into the circuit as shown in step C in Fig.~\ref{fig:interference}. Step D in the circuit loads the second training vector using controlled rotations. Loading training vectors of this specialized case resulted in a circuit with relatively shorter depth considering that we do not require controlled rotation to load  training vector $x^0=(0,1)$.

\begin{figure}[h]
\centering
\begin{subfigure}{1\columnwidth}
    \includegraphics[width=1\columnwidth]{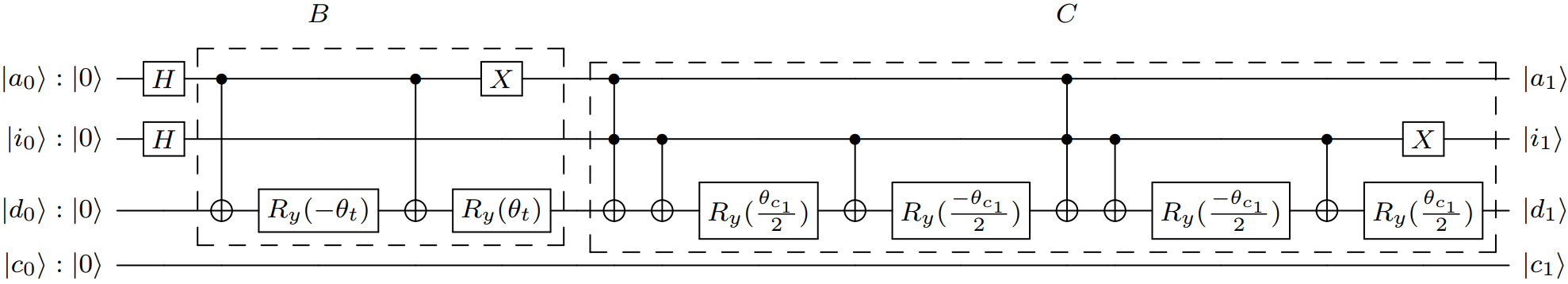}
\end{subfigure} \vspace{1cm}

\begin{subfigure}{1\columnwidth}
   \includegraphics[width=1\columnwidth]{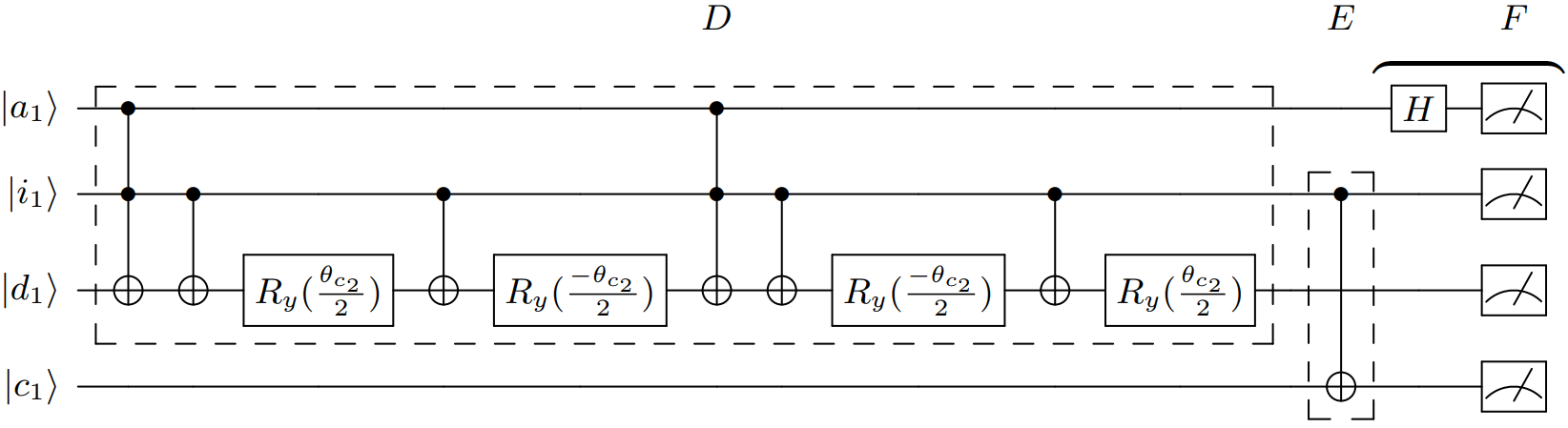}
\end{subfigure}

\caption{Distance Based Classifier Circuit with arbitrary training vectors \textbf{Step B:} Load test vector; \textbf{Step C:} Load training vector 1; \textbf{Step D:} Load training vector 2; \textbf{Step E:} Flip the class label for training vector 2; \textbf{Step F:} Interference and measurement.}
\label{fig:interference_modified}

\end{figure}

In order to adapt this approach to implement quantum K-means clustering algorithm we would need a circuit that can load arbitrary training vectors. In order to do that, let us consider a simplified K-means algorithm with $K = 2$ for only 2 clusters and training vectors having 2 dimensional data. Further, we will consider training vectors as centroids and since after each iteration of the K-means algorithm a new centroid is calculated, we will need to load arbitrary training vectors (centroids) into the circuit after each iteration. This modified circuit can load arbitrary training vectors with controlled rotations for both training vectors and it is shown in Fig.~\ref{fig:interference_modified}.

\subsubsection{Optimized 4 qubit model} \label{sec:4_qubit_model}
\begin{figure}[h]
    \centering
    \includegraphics[width=1\columnwidth]{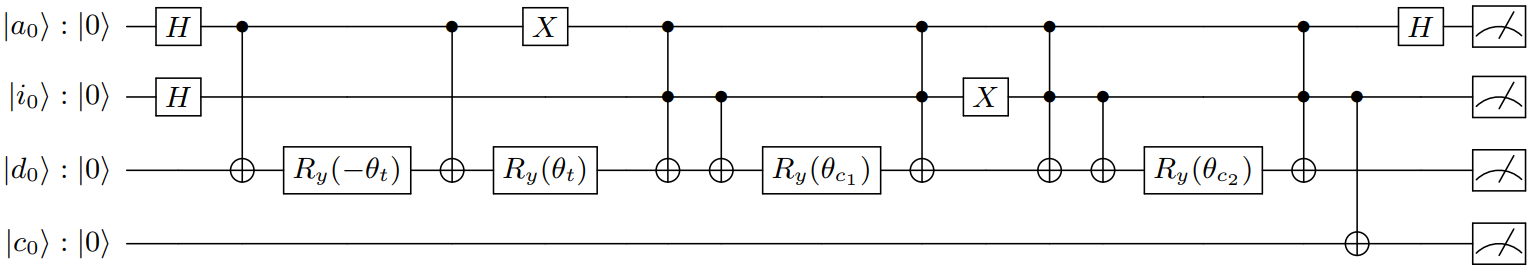}
    \caption{Optimized Interference Circuit.}
    \label{fig:optimized_interference}
\end{figure}

In this strategy we use the same number of qubits as in \cite{distanceclassifier:2017:maria} while reducing the depth of the quantum circuit. The desired quantum state required by the quantum interference to perform the classification is given by:

\begingroup
\fontsize{8pt}{8pt}\selectfont
\begin{equation}
  \ket{\psi} = \frac{1}{\sqrt{2}} \left( \ket{0} \ket{a} + \ket{1} \ket{b}  \right)
\label{eq:str1.2}
\end{equation}
\endgroup

where $\ket{a}$ is the quantum state for of the test vector and $\ket{b}$ is the quantum state of the training vectors. A Hadamard gate applied to the ancilla qubit of this quantum state performs the interference between the test $\ket{a}$ and training vectors $\ket{b}$. This is the desired quantum state where  $\ket{a}$ is entangled with the $\ket{0}$ state of ancilla qubit and $\ket{b}$ is entangled with the $\ket{1}$ state of the ancilla qubit. Another valid quantum state that would provide us with the same interference probability pattern can be written as follows:

\begingroup
\fontsize{8pt}{8pt}\selectfont
\begin{equation}
\ket{\psi_1} = \frac{1}{\sqrt{2}} \left( \ket{0} \ket{a_1} + \ket{1} \ket{b_1}  \right)
\label{eq:str1.3}
\end{equation}
\endgroup

which is similar to the state in \eqref{eq:str1.1} but here $\ket{a_1}$ and $\ket{b_1}$, the corresponding quantum states of test and training vectors respectively do not represent the input vectors instead they represent a new configuration of the input vectors. This new configuration is valid for quantum interference if the following equation holds true:

\begingroup
\fontsize{8pt}{8pt}\selectfont
\begin{equation}
|\theta_{\ket{a}} - \theta_{\ket{b}}| = |\theta_{\ket{a_1}} - \theta_{\ket{b_1}}|
\label{eq:str1.4}
\end{equation}
\endgroup

where $\theta$ denotes the angle of the respective vector. By preparing the state as shown in equation \eqref{eq:str1.3} we can load a new configuration of the test and training vectors and this quantum state can be reached with a shorter circuit depth. The resulting optimized circuit is shown in Fig.~\ref{fig:optimized_interference}. 

\begin{figure*}[h]
\centering
\begin{subfigure}{0.15\textwidth}
{\scalebox{0.8}{\begin{tikzpicture}[>=latex]
\draw[step=0.5cm,gray!25!,very thin] (-1.5,-1.5) grid (1.5,1.5);
\draw[thick,->] (-1.5,0) -- (1.5,0) node[anchor=north west] {};
\draw[thick,->] (0,-1.5) -- (0,1.5) node[anchor=south east] {};

\draw[thick,->] (0,0) coordinate (O) -- (45:1) coordinate (oc) 
node[midway,above] {$\vec{t}$};

\draw[blue,thick,->] (0,0) coordinate (O) -- (0:1) coordinate (oc) 
node[midway,below] {$\vec{c_1}$};

\draw[black!40,thick,->] (0,0) coordinate (O) -- (180:1) coordinate (oc) 
node[midway,below] {$\vec{c_2}$};

\draw[red, thick] (O) circle (1 cm);

\draw (0.25,0) arc (0:45:0.25);
\node[] at (20:0.5)  {$\theta_1$};

\draw (45:0.30) arc (45:180:0.25);
\node[] at (120:0.5)  {$\theta_2$};


\end{tikzpicture}}
\caption{}
\label{fig:example1a}}
\end{subfigure}  
\begin{subfigure}{0.15\textwidth}
{\scalebox{0.8}{\begin{tikzpicture}[>=latex]
\draw[step=0.5cm,gray!25!,very thin] (-1.5,-1.5) grid (1.5,1.5);
\draw[thick,->] (-1.5,0) -- (1.5,0) node[anchor=north west] {};
\draw[thick,->] (0,-1.5) -- (0,1.5) node[anchor=south east] {};

\draw[thick,->] (0,0) coordinate (O) -- (45:1) coordinate (oc) 
node[midway,above] {$\vec{t}$};

\draw[red, thick] (O) circle (1 cm);

\draw[->,semithick] (1.25,0) arc (0:45:1.25);
\node[] at (20:1.7)  {$R_y(\theta_t)$};


\end{tikzpicture}}
\caption{}
\label{fig:example3a}}
\end{subfigure}
\begin{subfigure}{0.15\textwidth}
{\scalebox{0.8}{\begin{tikzpicture}[>=latex]
\draw[step=0.5cm,gray!25!,very thin] (-1.5,-1.5) grid (1.5,1.5);
\draw[thick,->] (-1.5,0) -- (1.5,0) node[anchor=north west] {};
\draw[thick,->] (0,-1.5) -- (0,1.5) node[anchor=south east] {};

\draw[thick,->] (0,0) coordinate (O) -- (45:1) coordinate (oc) 
node[midway,above] {$\vec{t}$};

\draw[red, thick] (O) circle (1 cm);

\node[] at (45:1.3)  {$X$};


\end{tikzpicture}}
\caption{}
\label{fig:example3b}}
\end{subfigure}
\begin{subfigure}{0.15\textwidth}
{\scalebox{0.8}{\begin{tikzpicture}[>=latex]
\draw[step=0.5cm,gray!25!,very thin] (-1.5,-1.5) grid (1.5,1.5);
\draw[thick,->] (-1.5,0) -- (1.5,0) node[anchor=north west] {};
\draw[thick,->] (0,-1.5) -- (0,1.5) node[anchor=south east] {};

\draw[thick,->] (0,0) coordinate (O) -- (45:1) coordinate (oc) 
node[midway,above] {$\vec{t}$};

\draw[red, thick] (O) circle (1 cm);

\node[] at (45:1.7)  {$R_y\left(\frac{\theta_{c_1}}{2}\right)$};


\end{tikzpicture}}
\caption{}
\label{fig:example3c}}
\end{subfigure}
\begin{subfigure}{0.15\textwidth}
{\scalebox{0.8}{\begin{tikzpicture}[>=latex]
\draw[step=0.5cm,gray!25!,very thin] (-1.5,-1.5) grid (1.5,1.5);
\draw[thick,->] (-1.5,0) -- (1.5,0) node[anchor=north west] {};
\draw[thick,->] (0,-1.5) -- (0,1.5) node[anchor=south east] {};

\draw[thick,->] (0,0) coordinate (O) -- (135:1) coordinate (oc) 
node[midway,above] {$\vec{t}$};

\draw[red, thick] (O) circle (1 cm);

\draw[->,semithick] (45:1.25) arc (45:135:1.25);
\node[] at (45:1.7)  {$R_y\left(\frac{\theta_{c_2}}{2}\right)$};


\end{tikzpicture}}
\caption{}
\label{fig:example3d}}
\end{subfigure}
\begin{subfigure}{0.15\textwidth}
{\scalebox{0.8}{\begin{tikzpicture}[>=latex]
\draw[step=0.5cm,gray!25!,very thin] (-1.5,-1.5) grid (1.5,1.5);
\draw[thick,->] (-1.5,0) -- (1.5,0) node[anchor=north west] {};
\draw[thick,->] (0,-1.5) -- (0,1.5) node[anchor=south east] {};

\draw[thick,->] (0,0) coordinate (O) -- (45:1) coordinate (oc) 
node[midway,above] {$\vec{t'}$};

\draw[red, thick] (O) circle (1 cm);

\draw[->,semithick] (1.25,0) arc (0:45:1.25);
\node[] at (20:1.7)  {$R_y(\theta_t)$};


\end{tikzpicture}}
\caption{}
\label{fig:example5a}}
\end{subfigure}
\begin{subfigure}{0.15\textwidth}
{\scalebox{0.8}{\begin{tikzpicture}[>=latex]
\draw[step=0.5cm,gray!25!,very thin] (-1.5,-1.5) grid (1.5,1.5);
\draw[thick,->] (-1.5,0) -- (1.5,0) node[anchor=north west] {};
\draw[thick,->] (0,-1.5) -- (0,1.5) node[anchor=south east] {};

\draw[thick,->] (0,0) coordinate (O) -- (45:1) coordinate (oc) 
node[midway,above] {$\vec{t'}$};

\draw[red, thick] (O) circle (1 cm);

\node[] at (45:1.7)  {$R_y\left(\frac{\theta_{c_1}}{2}\right)$};


\end{tikzpicture}}
\caption{}
\label{fig:example5b}}
\end{subfigure}
\begin{subfigure}{0.15\textwidth}
{\scalebox{0.8}{\begin{tikzpicture}[>=latex]
\draw[step=0.5cm,gray!25!,very thin] (-1.5,-1.5) grid (1.5,1.5);
\draw[thick,->] (-1.5,0) -- (1.5,0) node[anchor=north west] {};
\draw[thick,->] (0,-1.5) -- (0,1.5) node[anchor=south east] {};

\draw[thick,->] (0,0) coordinate (O) -- (45:1) coordinate (oc) 
node[midway,above] {$\vec{t'}$};

\draw[red, thick] (O) circle (1 cm);

\node[] at (45:1.3)  {$X$};


\end{tikzpicture}}
\caption{}
\label{fig:example5c}}
\end{subfigure}
\begin{subfigure}{0.15\textwidth}
{\scalebox{0.8}{\begin{tikzpicture}[>=latex]
\draw[step=0.5cm,gray!25!,very thin] (-1.5,-1.5) grid (1.5,1.5);
\draw[thick,->] (-1.5,0) -- (1.5,0) node[anchor=north west] {};
\draw[thick,->] (0,-1.5) -- (0,1.5) node[anchor=south east] {};

\draw[thick,->] (0,0) coordinate (O) -- (135:1) coordinate (oc) 
node[midway,above] {$\vec{t'}$};

\draw[red, thick] (O) circle (1 cm);

\draw[->,semithick] (45:1.25) arc (45:135:1.25);
\node[] at (45:1.7)  {$R_y\left(\frac{\theta_{c_2}}{2}\right)$};


\end{tikzpicture}}
\caption{}
\label{fig:example5d}}
\end{subfigure}
\begin{subfigure}{0.15\textwidth}
{\scalebox{0.8}{\begin{tikzpicture}[>=latex]
\draw[step=0.5cm,gray!25!,very thin] (-1.5,-1.5) grid (1.5,1.5);
\draw[thick,->] (-1.5,0) -- (1.5,0) node[anchor=north west] {};
\draw[thick,->] (0,-1.5) -- (0,1.5) node[anchor=south east] {};

\draw[blue,->] (0,0) coordinate (O) -- (90:1) coordinate (oc) 
node[midway,above] {};

\node[blue] at (70:0.5)  {$\vec{c_1}$};

\draw[red, thick] (O) circle (1 cm);

\draw[->,semithick] (1.25,0) arc (0:90:1.25);
\node[] at (45:1.7)  {$X$};


\end{tikzpicture}}
\caption{}
\label{fig:example4a}}
\end{subfigure}
\begin{subfigure}{0.15\textwidth}
{\scalebox{0.8}{\begin{tikzpicture}[>=latex]
\draw[step=0.5cm,gray!25!,very thin] (-1.5,-1.5) grid (1.5,1.5);
\draw[thick,->] (-1.5,0) -- (1.5,0) node[anchor=north west] {};
\draw[thick,->] (0,-1.5) -- (0,1.5) node[anchor=south east] {};

\draw[blue,->] (0,0) coordinate (O) -- (90:1) coordinate (oc) 
node[midway,above] {};

\node[blue] at (70:0.5)  {$\vec{c_1}$};

\draw[red, thick] (O) circle (1 cm);

\node[] at (45:1.7)  {$R_y\left(-\frac{\theta_{c_1}}{2}\right)$};


\end{tikzpicture}}
\caption{}
\label{fig:example4b}}
\end{subfigure}
\begin{subfigure}{0.15\textwidth}
{\scalebox{0.8}{\begin{tikzpicture}[>=latex]
\draw[step=0.5cm,gray!25!,very thin] (-1.5,-1.5) grid (1.5,1.5);
\draw[thick,->] (-1.5,0) -- (1.5,0) node[anchor=north west] {};
\draw[thick,->] (0,-1.5) -- (0,1.5) node[anchor=south east] {};

\draw[blue,->] (0,0) coordinate (O) -- (180:1) coordinate (oc) 
node[midway,above] {$\Vec{c_1}$};

\draw[red, thick] (O) circle (1 cm);

\draw[->,semithick] (90:1.25) arc (90:180:1.25);
\node[] at (45:1.7)  {$R_y\left(\frac{\theta_{c_2}}{2}\right)$};


\end{tikzpicture}}
\caption{}
\label{fig:example4c}}
\end{subfigure}    
\begin{subfigure}{0.15\textwidth}
{\scalebox{0.8}{\begin{tikzpicture}[>=latex]
\draw[step=0.5cm,gray!25!,very thin] (-1.5,-1.5) grid (1.5,1.5);
\draw[thick,->] (-1.5,0) -- (1.5,0) node[anchor=north west] {};
\draw[thick,->] (0,-1.5) -- (0,1.5) node[anchor=south east] {};

\draw[black!50,->] (0,0) coordinate (O) -- (0:1) coordinate (oc) 
node[midway,above] {$\Vec{c_2}$};

\draw[red, thick] (O) circle (1 cm);

\node[] at (45:1.7)  {$R_y\left(\frac{\theta_{c_1}}{2}\right)$};


\end{tikzpicture}}
\caption{}
\label{fig:example6a}}
\end{subfigure}
\begin{subfigure}{0.15\textwidth}
{\scalebox{0.8}{\begin{tikzpicture}[>=latex]
\draw[step=0.5cm,gray!25!,very thin] (-1.5,-1.5) grid (1.5,1.5);
\draw[thick,->] (-1.5,0) -- (1.5,0) node[anchor=north west] {};
\draw[thick,->] (0,-1.5) -- (0,1.5) node[anchor=south east] {};

\draw[black!50,->] (0,0) coordinate (O) -- (90:1) coordinate (oc) 
node[midway,above] {};

\node[black!50] at (70:0.5)  {$\vec{c_2}$};

\draw[red, thick] (O) circle (1 cm);

\draw[->,semithick] (1.25,0) arc (0:90:1.25);
\node[] at (45:1.7)  {$X$};


\end{tikzpicture}}
\caption{}
\label{fig:example6b}}
\end{subfigure}
\begin{subfigure}{0.15\textwidth}
{\scalebox{0.8}{\begin{tikzpicture}[>=latex]
\draw[step=0.5cm,gray!25!,very thin] (-1.5,-1.5) grid (1.5,1.5);
\draw[thick,->] (-1.5,0) -- (1.5,0) node[anchor=north west] {};
\draw[thick,->] (0,-1.5) -- (0,1.5) node[anchor=south east] {};

\draw[black!50,->] (0,0) coordinate (O) -- (0:1) coordinate (oc) 
node[midway,above] {$\Vec{c_2}$};

\draw[red, thick] (O) circle (1 cm);

\draw[->,semithick] (90:1.25) arc (90:0:1.25);
\node[] at (45:1.8)  {$R_y\left(-\frac{\theta_{c_2}}{2}\right)$};


\end{tikzpicture}}
\caption{}
\label{fig:example6c}}
\end{subfigure}
\begin{subfigure}{0.15\textwidth}
{\scalebox{0.8}{\begin{tikzpicture}[>=latex]
\draw[step=0.5cm,gray!25!,very thin] (-1.5,-1.5) grid (1.5,1.5);
\draw[thick,->] (-1.5,0) -- (1.5,0) node[anchor=north west] {};
\draw[thick,->] (0,-1.5) -- (0,1.5) node[anchor=south east] {};

\draw[thick,->] (0,0) coordinate (O) -- (135:1) coordinate (oc) 
node[midway,above] {$\vec{t}$};

\draw[blue,thick,->] (0,0) coordinate (O) -- (180:1) coordinate (oc) 
node[midway,below] {$\vec{c_1}$};

\draw[red, thick] (O) circle (1 cm);

\draw (135:0.3) arc (135:180:0.25);
\node[] at (160:0.5)  {\small $\theta_3$};


\end{tikzpicture}}
\caption{}
\label{fig:example7}}
\end{subfigure}
\begin{subfigure}{0.15\textwidth}
{\scalebox{0.8}{\begin{tikzpicture}[>=latex]
\draw[step=0.5cm,gray!25!,very thin] (-1.5,-1.5) grid (1.5,1.5);
\draw[thick,->] (-1.5,0) -- (1.5,0) node[anchor=north west] {};
\draw[thick,->] (0,-1.5) -- (0,1.5) node[anchor=south east] {};

\draw[thick,->] (0,0) coordinate (O) -- (135:1) coordinate (oc) 
node[midway,above] {$\vec{t'}$};

\draw[black!50,thick,->] (0,0) coordinate (O) -- (0:1) coordinate (oc) 
node[midway,below] {$\vec{c_2}$};

\draw[red, thick] (O) circle (1 cm);

\draw (0:0.25) arc (0:135:0.25);
\node[] at (45:0.4)  {$\theta_4$};


\end{tikzpicture}}
\caption{}
\label{fig:example8}}
\end{subfigure}
\begin{subfigure}{0.15\textwidth}
{\scalebox{0.8}{\begin{tikzpicture}[>=latex]
\draw[step=0.5cm,gray!25!,very thin] (-1.5,-1.5) grid (1.5,1.5);
\draw[thick,->] (-1.5,0) -- (1.5,0) node[anchor=north west] {};
\draw[thick,->] (0,-1.5) -- (0,1.5) node[anchor=south east] {};

\draw[thick,->] (0,0) coordinate (O) -- (135:1) coordinate (oc) 
node[midway,above] {$\vec{t}$};

\draw[black!40,thick,->] (0,0) coordinate (O) -- (0:1) coordinate (oc) 
node[midway,below] {$\vec{c_2}$};

\draw[blue,thick,->] (0,0) coordinate (O) -- (180:1) coordinate (oc) 
node[midway,below] {$\vec{c_1}$};

\draw[red, thick] (O) circle (1 cm);

\draw (135:0.25) arc (135:180:0.25);
\node[] at (160:0.5)  {$\theta_3$};

\draw (0:0.25) arc (0:135:0.30);
\node[] at (45:0.5)  {$\theta_4$};


\end{tikzpicture}}
\caption{}
\label{fig:example1b}}
\end{subfigure}
\caption{[b-e]: Quantum state transition of vector $\Vec{t}$ from equation~\eqref{eq:t}, [f-i]: Quantum state transition of vector $\Vec{t'}$ from equation~\eqref{eq:tcopy}, [j-l]: Quantum state transition of vector $\vec{c_1}$ from equation~\eqref{eq:c1}, [m-o]: Quantum state transition of vector $\vec{c_2}$ from equation~\eqref{eq:c2}, [p-q]: Final State of interfering vectors [r]: Combined final state of the interfering vectors}
\label{fig:example7and8}
\end{figure*}
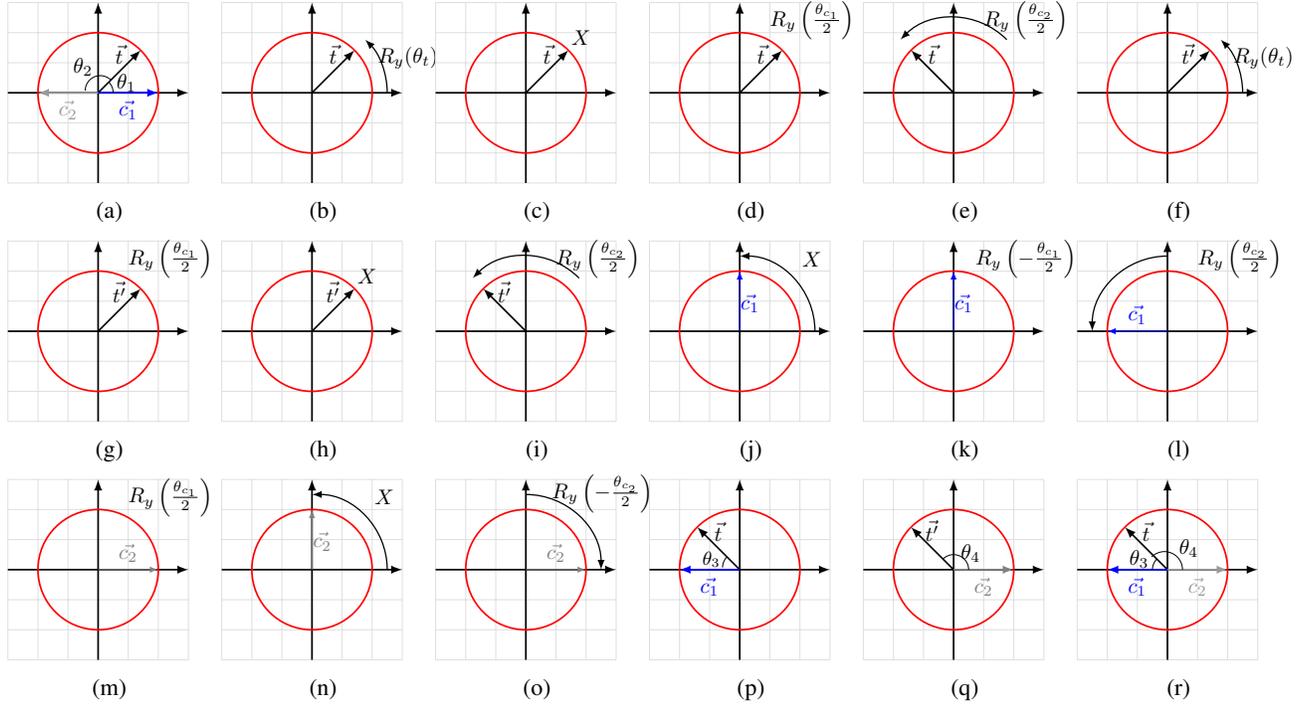

Fig.~\ref{fig:optimized_interference} shows the optimized distance based classifier circuit with arbitrary training vectors. The optimization is based on achieving the same relative angular difference between training and test vectors as the circuit in~\cite{distanceclassifier:2017:maria} with fewer quantum gates.
To elaborate it further, let us take an example of one test and two training vectors, and let us assume that the training vector 1 has a $0^\circ$ rotation from the x axis, the training vector 2 has a $180^\circ$ rotation, and test vector has $45^\circ$ rotation. The angle between the training vector 1 and test vector is $45^\circ$ while the angle between the training vector 2 and the test vector is $135^\circ$ as shown in Fig.~\ref{fig:example1a}. This is how the original circuit would load the test and the training vectors in the entangled quantum states. Note that, in order to obtain the same interference pattern at the end of the quantum circuit we do not necessarily need to load our vectors into the quantum states that directly correspond to their input states. We will get the same interference pattern as long as the angle between the test and training vectors is the same at the end of the state preparation. For instance, Fig.~\ref{fig:example1b} shows an example of a valid configuration that adheres to the constraint~\eqref{eq:str1.3}. Both of these configurations shown in Fig.~\ref{fig:example1a} and Fig.~\ref{fig:example1b} yield to the same interference pattern probabilities after measurement.

We exploit this property of interference pattern probabilities to reduce the depth of the circuit. The state preparation part of the optimized interference circuit is not constrained to load the exact states of the input vectors. It is designed to load the interfering copies of the input vectors with the same relative angular difference.

The goal of loading the vectors with same relative angular difference is reached by making the input vectors undergo a certain set of quantum operations within the state preparation phase. The operations that are performed on each of the input vectors are outlined below. Note that the initial state of each vector is  $\bigl( \begin{smallmatrix}1\\0\end{smallmatrix}\bigr)$,

\begingroup
\fontsize{8pt}{8pt}\selectfont
\begin{equation} 
t: R_y\left(\frac{\theta_{c_2}}{2}\right) R_y\left(\frac{\theta_{c_1}}{2}\right) \  X \  R_y(\theta_t) \begin{pmatrix} 1\\0\end{pmatrix}
\label{eq:t}
\end{equation}

\begin{equation}
t': R_y\left(\frac{\theta_{c_2}}{2}\right)
X \ 
R_y\left(\frac{\theta_{c_1}}{2}\right) \ 
R_y(\theta_t) \ 
\begin{pmatrix} 1\\0\end{pmatrix}
\label{eq:tcopy}
\end{equation}
 
\begin{equation}
c_1: R_y\left(\frac{\theta_{c_2}}{2}\right)
R_y\left(-\frac{\theta_{c_1}}{2}\right) \ 
X \ 
\begin{pmatrix} 1\\0\end{pmatrix}
\label{eq:c1}
\end{equation}

\begin{equation}
c_2: R_y\left(-\frac{\theta_{c_2}}{2}\right)
X \ 
R_y\left(\frac{\theta_{c_1}}{2}\right) \ 
\begin{pmatrix} 1\\0\end{pmatrix}
\label{eq:c2}
\end{equation}
\endgroup

Here $t$ is the test vector and $t'$ is its copy. $c_1$ and $c_2$ are two training vectors. $X$ is a simple quantum $NOT$ operation and $R_y$ represents the rotation operation. In this case, $t$ interferes with $c_1$ and $t'$ interferes with $c_2$. Hence, the angular difference between $t$ and $c_1$: $\theta_{\theta_t - \theta_{c_1}}$ and the angle between $t'$ and $c_2$: $\theta_{\theta_{t'} - \theta_{c_2}}$ in the original configuration should be equal to the corresponding angles in the new  configuration that comes from the optimized circuit. An example of the quantum state evolution of the test and training vectors is shown in Fig.~\ref{fig:example3a} to Fig.~\ref{fig:example6c}.

Fig.~\ref{fig:example1a} shows the original configuration of our example. We have two training vectors $\vec{c_1}$ and $\vec{c_2}$ and one test vector  $\vec{t}$ along with the associated angular difference between the interfering vectors: $\theta_1$ and $\theta_2$. Fig.~\ref{fig:example1b} shows the final state of the interfering vectors at the end of the state preparation phase of the optimized quantum interference circuit. As it can be seen from Fig.~\ref{fig:example1a} and~\ref{fig:example1b}, $\theta_1$ is equal to the corresponding $\theta_3$, and $\theta_2$ is equal to $\theta_4$ providing us with the required result, in agreement with eq. \eqref{eq:str1.3}.

The optimized circuit can also be used in the state preparation phase for any algorithm where the relative rotation of vectors is of account instead of the input states. However,  it is important to mention that this optimization is specifically designed for classification/clustering problems where only the relative angular rotation is accounted and uses interference as a metric for final decision. This will not work where exact input states are required to be loaded in the quantum circuit.

\subsubsection{Multi-cluster Quantum K-means} \label{sec:multi_cluster}
\begin{figure}[ht]
    \centering
    \includegraphics[scale=0.35]{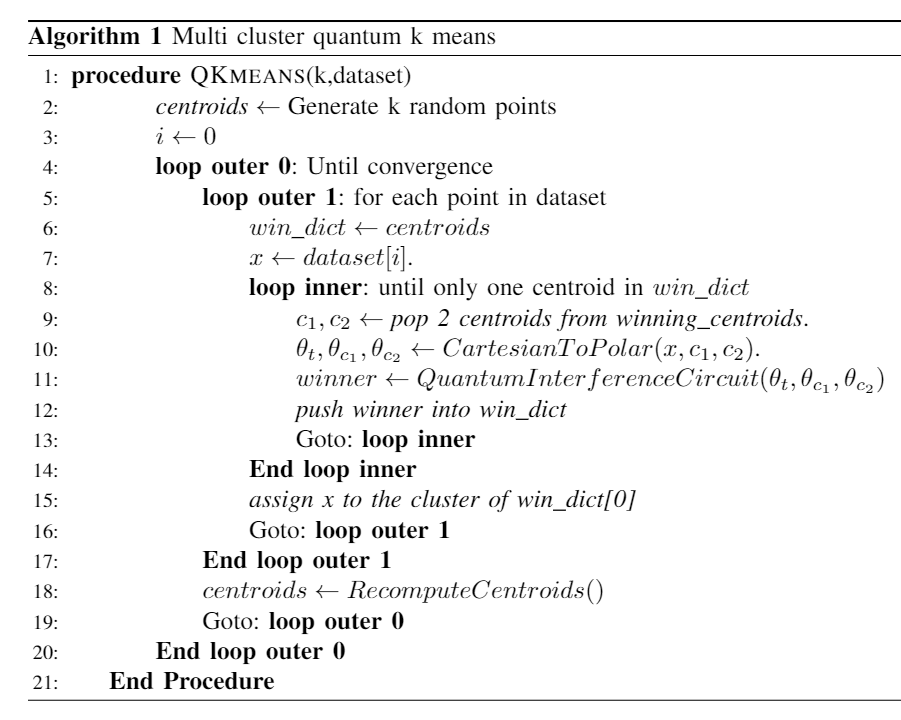}
    \caption{Multi cluster quantum K-means}
    \label{fig:kmeans}
\end{figure}

The quantum K-means implementation that we have discussed in the previous section can only group data into two clusters. Here, we propose a strategy to perform multi cluster ($K>2$) quantum K-means using a constant size circuit. In this strategy we use an elimination method to find the nearest centroid to the test vector. The approach is outlined in Fig.~\ref{fig:kmeans}.

The input to the algorithm is K (number of clusters) and a set of input vectors. The first step of the algorithm is to generate K centroids randomly. After this we run the outer loop which goes through all input vectors and assigns clusters to them. Inside this outer loop, centroids compete (two at a time) with each other to acquire the current input vector into their cluster. In each round, the centroid which is closer to the input vector wins and takes part in the next round. The classical computer maintains a dictionary $win\_dict$ that stores all the centroids that won the last round or centroids that are yet to take their turn. $win\_dict$ is initialized with K centroids. Two centroids ($c_i, c_j$) are taken out from this dictionary, and along with the input vector ($t_i$) they are loaded into the quantum interference circuit. The quantum interference circuit is executed on the quantum computer and interference probabilities are returned to the classical computer that filters out the constructive interference probabilities. The nearer centroid is found based on the probability of the least significant qubit. If the probability of state $\ket{0}$ is high then $c_i$ wins, else $c_j$ wins. Win in this context means the centroid is closer to the input vector. The winning centroid is then pushed back into the $win\_dict$. This process is repeated until we have only one remaining centroid in $win\_dict$. The input vector is assigned to the cluster of that remaining centroid. The whole process is repeated for all the vectors in the input dataset. After assigning clusters to each input vector, centroids are recomputed by averaging all vectors in respective clusters. These new centroids are then used to do the cluster assignment again for all the vectors. This procedure is repeated until there is no change in cluster assignments in subsequent iterations.

\subsection{Constant Depth circuit using Negative Rotations}  \label{sec:negative_rotations}
Here we propose a second solution to perform clustering using a constant depth circuit. This implementation uses the negative (or opposite) rotations of the qubits to find the nearest centroid to the test vector. Given a test vector $v$ and two centroids $\{c_1 , c_2\}$ we want to find out which of the two centroids is closer to $v$. The closest centroid would have a smaller angular difference, so if $|\theta_t - \theta_{c_1}|$ is less than $|\theta_t - \theta_{c_2}|$ then $t$ is closer to $c_1$ else is closer to $c_2$.

In terms of implementing this on a quantum computer using qubits we need to take into account the probabilities of a qubit being in state $\ket{0}$ and state $\ket{1}$. Knowing that the probability of a qubit being in state $\ket{0}$ is higher for those vectors close to x-axis, we will use this as a metric to find which centroid lies closest. We want the probability of the qubit in state $\ket{0}$ to be higher for those vectors that are close and the probability of state $\ket{1}$ to be higher when they are further away. The maximum angle between two vectors is $180^o$, meaning when the vectors are at max distance the angle between them is $180^o$

\[ |\theta_t - \theta_c| \leq 180^o  \]

In this manner, a qubit rotated at $Ry(180^o)$ has $100\%$ probability to be in state $\ket{0}$, however we want the probability of state $\ket{1}$ to be higher when the vectors are $180^o$ apart. Therefore, in this implementation we will use the qubit rotated to $Ry(90^o)$ to define the max angle between two vectors. This can be achieved by doing half rotations, 

\[ | \frac{\theta_t}{2}  - \frac{\theta_c}{2}| \leq \frac{180^o}{2}  \]

\begin{figure}[h]
\makebox[1.1\linewidth]{
\input{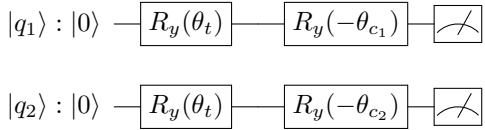}}
\caption{Two qubit Negative Rotations Quantum Circuit.}
\label{fig:netative_rotation_2}
\end{figure}

The quantum circuit implementing negative rotations for a problem with two centroids is shown in Fig.~\ref{fig:netative_rotation_2}.

This is a two step process. The first step is to rotate the qubit to $\frac{\theta_t}{2}$, note that $Ry(\theta_t)$ does half rotation (see section \ref{sec:quantum_gates}). The second step is to do the negative rotation: $-\frac{\theta_{c_1}}{2}$ and $-\frac{\theta_{c_2}}{2}$ on the respective qubits. Finally, the measurement is done on both qubits and the probability of state $\ket{0}$ is compared. Cluster assignment of the test vector $t$ is done when the probability of the qubit in state $\ket{0}$ is high. In other words, if the probability of state $\ket{0}$ for $q_1$ is higher than $q_2$ then $c_1$ is closer to $t$ than $c_2$. 

\begin{figure}[h]
    \centering
    \includegraphics[width=1\columnwidth]{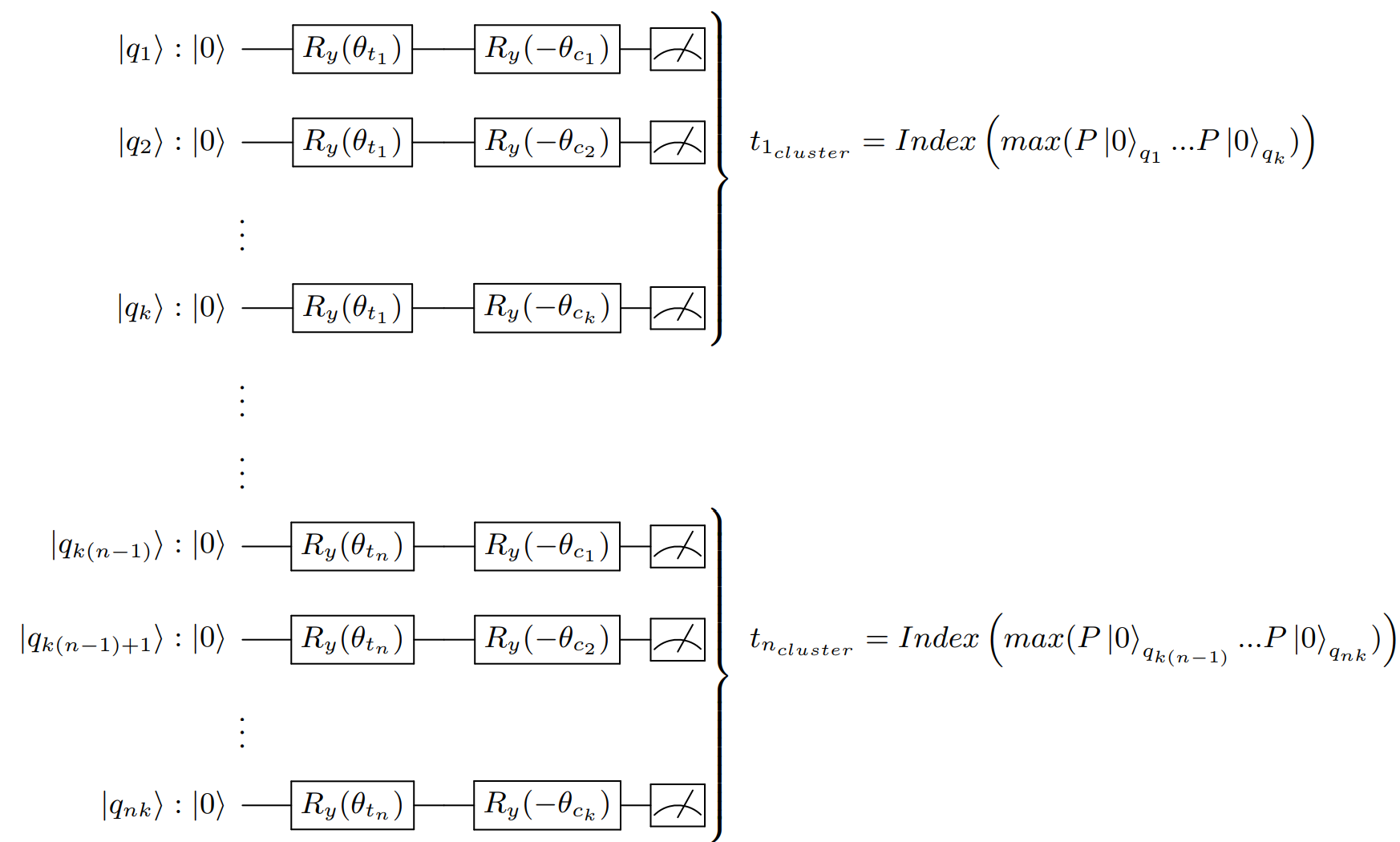}
    \caption{$nk$ qubit negative rotation circuit.}
    \label{fig:negative_rotation_nk}
\end{figure}

A generalized circuit for $K$ clusters and $n$ test vectors is shown in Fig. \ref{fig:negative_rotation_nk}. In this implementation, if we want to cluster $n$ test vectors in $K$ clusters we need $nk$ qubits. The depth of the circuit remains constant, however the width (number of qubits) depends on the number of test vectors and clusters.

\subsection{Distance Calculation using Destructive Interference} \label{sec:destructive_interference}
Here we propose an alternative way of calculating distances which is used to implement quantum K-means algorithm. This approach uses destructive interference probabilities to calculate distances between vectors. These distances are then used to find the nearest centroid.

\begin{figure}[h]
    \centering
    \scalebox{0.75}{
    \input{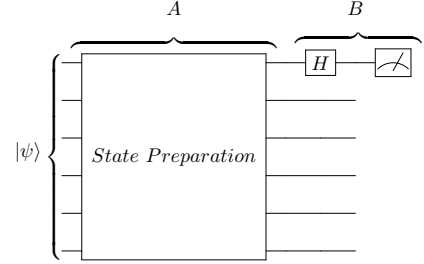}
    }
    \caption{Generic $n$ qubit Quantum Interference Circuit. \textbf{Step A:} Prepare the desired state; \textbf{Step B:} Interference and Measurement.}
    \label{fig:generic_interference}
\end{figure}

There are two main stages of a quantum interference circuit: State Preparation and Interference. Fig.~\ref{fig:generic_interference} shows a generic model of $n$ qubit Quantum Interference Circuit. $\ket{\psi}$ represents an $n$ qubit quantum state prepared in phase A. Once the desired quantum state is prepared, we just need to use the Hadamard gate (phase B) to perform interference on our desired state.

The distance calculation between two vectors to perform the Quantum K-means clustering is done using only the destructive interference probabilities. Given two vectors $t=[t_x, t_y]$ and $c=[c_x, c_y]$, the Euclidean distance between these two vectors is calculated as, 

\begingroup
\fontsize{8pt}{8pt}\selectfont
\[ d(t,c) = \sqrt{(t_x-c_x)^2 + (t_y-c_y)^2}\]
\endgroup

Recall from section \ref{sec:quantum_gates} that applying a Hadamard gate to a two qubit quantum state, we get destructive interference probabilities at $\ket{1}$ state of the most significant qubit. 

\begingroup
\fontsize{8pt}{8pt}\selectfont
\[\ket{\psi} = \begin{bmatrix}t'_x&t'_y&c'_x&c'_y \end{bmatrix}^T \]
\endgroup

here $\ket{\psi}$ is a normalized quantum state of the original vectors

\begingroup
\fontsize{8pt}{8pt}\selectfont
\[Norm = \sqrt{t_x^2+t_y^2 + c_x^2+c_y^2}\]
\[t'_x = \frac{t_x}{Norm}, t'_y = \frac{t_y}{Norm}, c'_x = \frac{c_x}{Norm},c'_y = \frac{c_y}{Norm} \]

\[ \ket{\phi} = \frac{1}{\sqrt{2}} \begin{pmatrix}1&0&1&0\\0&1&0&1\\1&0&-1&0\\0&1&0&-1\end{pmatrix} \begin{pmatrix}t'_x\\t'_y\\c'_x\\c'_y \end{pmatrix} = \frac{1}{\sqrt{2}} \begin{pmatrix}t'_x+c'_x\\t'_y+c'_y\\t'_x-c'_x\\t'_y-c'_y \end{pmatrix} \begin{matrix}\ket{00}\\\ket{01}\\\ket{10}\\\ket{11} \end{matrix}  \]
\endgroup

The probability of the most significant qubit in $\ket{\phi}$ to be in state $\ket{1}$ is given by,

\begingroup
\fontsize{8pt}{8pt}\selectfont
\[P\ket{1} =  \frac{1}{2} \left[  (t'_x - c'_x)^2 + (t'_y - c'_y)^2 \right]\]
\endgroup

we can use this probability to easily show that the actual distance between $t$ and $c$ is,

\begingroup
\fontsize{8pt}{8pt}\selectfont
\[ d(t,c) = Norm \times \sqrt{2} \sqrt{P\ket{1}}   \]
\endgroup

\begin{figure}
    \centering
    \includegraphics[width=1\columnwidth]{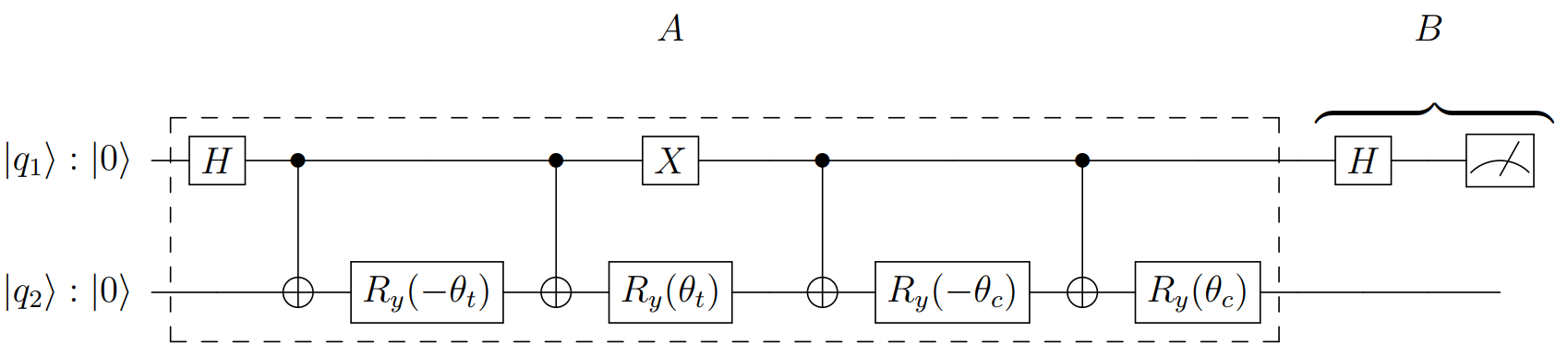}
    \caption{Two qubit quantum interference circuit.  \textbf{Step A:} Prepare the desired state; \textbf{Step B:} Interference and Measurement.}
    \label{fig:2qubit_distance}
\end{figure}

A two qubit quantum interference circuit is designed to implement this solution. The circuit is given in Fig.~\ref{fig:2qubit_distance}. Vector $t$ is entangled with the $\ket{0}$ state of qubit $\ket{q_1}$ and vector $c$ is entangled with $\ket{1}$ state of the qubit $\ket{q_1}$. Interference is done at the end and probabilities are measured only where the qubit $\ket{q_1}$ is in state $\ket{1}$ (to get destructive interference probabilities).

The distance calculation part of the K-means algorithm is done using this approach. For each test vector $t$, $K$ distances $[d_1,d_2,...d_k]$ are calculated iteratively running the circuit, where $d_1$ is the distance $d(t,c_1)$. $t$ is assigned to the cluster of the closest centroid $min([d_1,d_2,...d_k])$.

The circuit shown in Fig. \ref{fig:2qubit_distance} prepares the exact quantum state of the two input vectors with two dimensions. Using our finding from section~\ref{sec:4_qubit_model}, that the probability pattern of interference remains unchanged as long as the interfering input vectors are loaded with equal relative angular difference, we can optimize the state preparation phase of our circuit.  

\begin{figure}
    \centering
    \includegraphics[width=1\columnwidth]{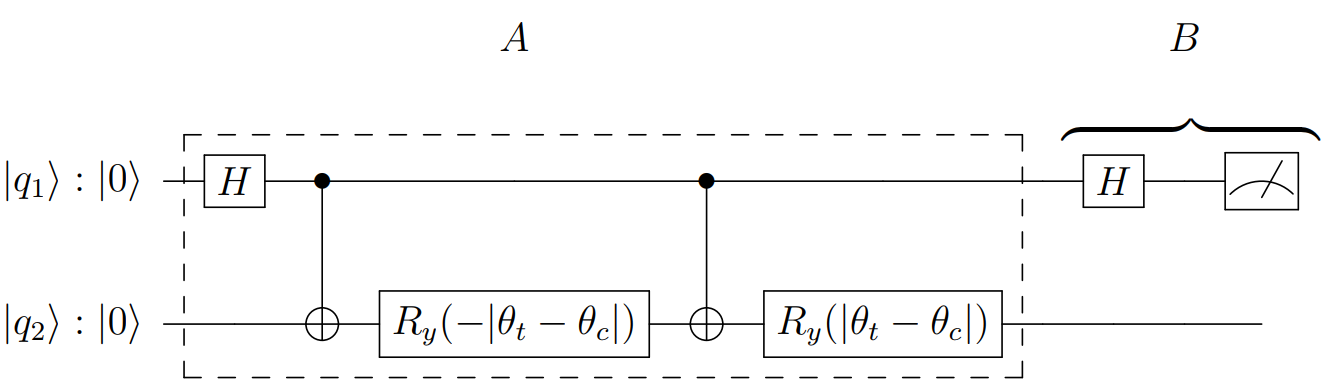}
    \caption{ Two qubit quantum interference circuit with equal relative angular difference. \textbf{Step A:} Prepare the desired state; \textbf{Step B:} Interference and Measurement.}
    \label{fig:2qubit_distance_optimized}
\end{figure}

Fig.~\ref{fig:2qubit_distance_optimized} shows a two qubit quantum interference circuit with equal relative angular difference. Vector $t$ is kept flat on the x-axis, vector $c$ is rotated with an angle $|\theta_t - \theta_c|$. This keeps both the vectors at the same angular difference as they were in the input state.

This approach is better than the $DistCal$ subroutine from section~\ref{sec:qkmeans} in terms of quantum operations and qubits required to calculate the distance. Here, we only need to prepare one state $\ket{\psi}$ unlike the $DistCal$ approach where two quantum states are required ($\ket{\psi}, \ket{\phi}$). Furthermore, $SwapTest$ subroutine is not required in this approach which reduces the quantum operations used in the circuit. Our proposed approach to calculate distance using destructive interference is much more efficient both in terms number of qubits required and total number of quantum operations. Given that the $Norm$ of the desired state is provided, besides preparing the desired quantum state, we only need one Hadamard gate and one measurement on the most significant qubit to calculate the distance between two vectors.

\section{Evaluation Methodology} \label{sec:evaluation}

\subsection{Input Data}
The implementations of the quantum K-means algorithm provided in this work are tested on three kinds of input datasets: Randomly Generated Dataset, Iris Dataset \cite{iris:1936} and MNIST Digits Dataset \cite{mnist-2010}.

\subsubsection{Random Dataset}  
100 input vectors with 2 dimensions are randomly generated with two dimensions. Each input vector is arbitrarily assigned to a cluster at the start. The dataset is then standardized to have zero mean and unit variance. This is commonly done in machine learning to avoid scaling effects and it is shown in Fig.~\ref{fig:random_dataset_2}. Next step of the pre-processing is to normalize each input vector in the input space to the unit length. This technique is also used in machine learning specially where only the angle between input vectors is required to distinguish between different classes. 
 
\subsubsection{Iris Dataset} 
The Iris dataset contains 50 samples from 3 different species of Iris flower. Each sample has 4 dimensions in total. We take only the first two dimensions from it which are sepal length and sepal width. There are three classes in the dataset that correspond to the names of the species: setosa, versicolor and virginica. So in this case, we hope to get three distinct clusters after running K-means algorithm. Here again we standardize and normalize the dataset. The dataset along with its true class labels is shown in Fig.~\ref{fig:iris_dataset}.

\subsubsection{MNIST Dataset} 
\begin{figure}[h]
\begin{subfigure}[c]{0.2\textwidth}
\includegraphics[scale=0.4]{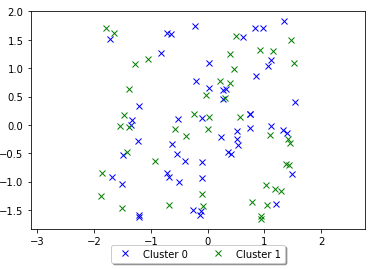}
\caption{}
\label{fig:random_dataset_2}
\end{subfigure}
\hskip2em
\begin{subfigure}[c]{0.2\textwidth}
\includegraphics[scale=0.4]{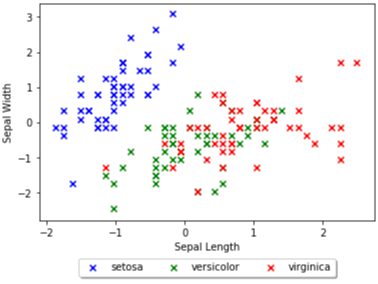}
\caption{}
\label{fig:iris_dataset}
\end{subfigure}

\begin{subfigure}[c]{0.2\textwidth}
\includegraphics[scale=0.4]{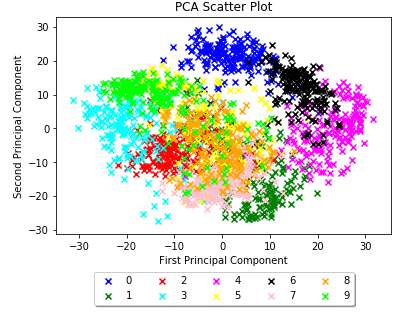}
\caption{}
\label{fig:mnist_pca_full}
\end{subfigure}
\hskip2em
\begin{subfigure}[c]{0.2\textwidth}
\includegraphics[scale=0.4]{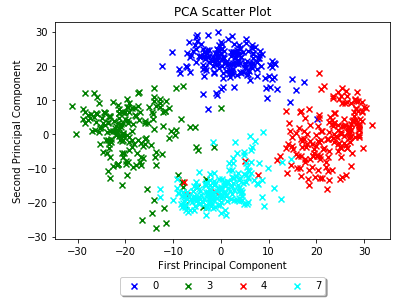}
\caption{}
\label{fig:mnist_pca_filtered}
\end{subfigure}
\caption{(a) Randomly Generated Dataset for K=2, (b) Iris Dataset; PCA reduced MNIST dataset with (c) true class labels, (d) true class labels containing digits 0,3,4,7}
\label{fig:mnist_pca}
\end{figure}

MNIST dataset contains grey scale images of handwritten digits. It is widely used in machine learning for training and testing different models. We use this dataset to test the clustering of digits using our implementations of the K-means algorithm. The images are represented as 28x28 matrices. That gives us a total of 784 features. In our implementation we only require two features. In this case,  we use a dimensionality reduction technique called Principal Component Analysis (PCA)  to reduce 784 features into 2 features. The PCA reduced MNIST dataset is shown in Fig.~\ref{fig:mnist_pca_full}. As shown from the PCA scatter plot, the clusters are not linearly separable. Therefore, we take only the digits that are separated in distinct clusters so that we can get meaningful results from our K-means implementation. Fig.~\ref{fig:mnist_pca_filtered} shows the PCA scatter plot for only four digits (0,3,4,7)

\subsection{Execution Environment}
The implementation of the solutions and their execution is done using IBM's open source quantum software development kit QISKIT \cite{qiskit}. QISKIT provides the framework to implement quantum circuits using python as the programming language. QISKIT allows us to run the quantum circuits on local quantum simulator and also on real quantum computers. Access to real quantum computers is provided through IBM Quantum Experience using an API access key.

Quantum circuits of the solutions presented in this work were executed first on the local quantum simulator and then on the real quantum computer (IBMQX2). Each run of the circuit was executed 8192 times to get a good estimate on the probabilities of the final quantum state.

\subsubsection{IBM Qiskit Simulator}
Qiskit provides access to multiple quantum simulators on a local computer as well as on the cloud. There is a unitary simulator that can simulate up to 12 qubits and a state vector simulator that can simulate circuits up to 25 qubits.

Running quantum circuits on the simulator provides ideal results as the simulator is noise free. There is a possibility of adding a noise model of choice into the simulator to make it mimic the actual quantum computer. However, for the executions of the circuits in this research work no noise was added into the simulator. Noise free simulators are used to make sure that if there are any inaccuracies in the results, that should be a consequence of incorrect circuit implementation and not of the noise in the simulator. The probabilities measured from the simulators are the ones that are theoretically predicted. This gives us the expected results from an actual noise free futuristic quantum computer. 

\subsubsection{IBMQX2 quantum machine}
IBMQX2 \cite{ibmq_devices} is a 5 superconducting qubit quantum computer. Running circuits on quantum computers from IBM quantum experience requires the user to push their job into a queue. After waiting for its turn the job is then executed on the quantum computer and results are returned back to the user. IBMQX2 specifications are shown in table \ref{tab:ibmqx2_specifications}. Due to low coherence times, gate errors and lack of error correction the results from IBMQX2 are expected to have errors depending on the depth of the circuit.

\begin{table}[h]
    \centering
    \resizebox{\columnwidth}{!}{
    \begin{tabular}{c|c|c|c} \hline
    \textbf{Qubit} & \textbf{T1($\mu s$) Relaxation Time}  & \textbf{T2($\mu s$) Coherence Time} & \textbf{Readout Error}  \\ \hline
    q0 & 57.3 & 47.4 & 0.08 \\ \hline
    q1 & 18.6 & 36.7 & 0.30  \\ \hline
    q2 & 75.9 & 68.3 & 0.01  \\ \hline
    q3 & 54.9 & 24.2 & 0.10 \\ \hline
    q4 & 50.8 & 70.9 & 0.39 \\ \hline
    \end{tabular}}
    \caption{IBMQX2 specifications}
    \label{tab:ibmqx2_specifications}
    
\end{table}

\subsection{Evaluation Metric}
Generally, the evaluation of unsupervised machine learning algorithms is non-trivial. Considering that the unlabelled data does not provide the ground truth of the true classes, calculating success rate of an unsupervised machine learning algorithm is not possible. One commonly used technique to compute accuracies of unsupervised/clustering machine learning models is to use labelled data. The evaluation of quantum K-means is done using this technique. The accuracy of the different implementations is calculated in 4 steps: i) take labelled dataset and remove labels from the input dataset; ii) execute the quantum K-means algorithm on the unlabelled dataset; iii) find the correlation between the final clusters and the true classes in the input dataset and; iv) use it to calculate the accuracy of the clustering algorithm.

In the case of the random dataset, results from the simulation are considered ideal,  and the accuracy of the result from ibmqx2 is calculated considering simulation results as true class labels. For Iris and MNIST datasets true class labels are given, which are used to calculate the accuracy. The quantum circuit is run 8192 $\times$ 10 times to get a good estimate on the performance of the algorithm.

\section{Results and discussion}\label{sec:results}
The solutions presented in section~\ref{sec:qkmeans_implementations} were implemented in QISKIT and executed first on simulator and then on IBMQX2. We also present results from classical K-means algorithm to compare it's performance with that of the quantum K-means algorithms. Scikit's~\cite{scikit-learn} version of the K-means algorithm was used for this purpose. However, a classical result for the random dataset is not used for comparison, because the random dataset K-means algorithm converges differently depending on the choice of initial centroids.

\subsection{Constant Depth Circuit using Interference}
\subsubsection{Basic Implementation}
\begin{figure}[h]
\begin{subfigure}{0.2\textwidth}
\includegraphics[scale=0.4]{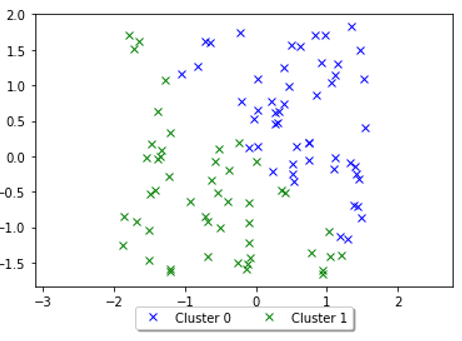}
\caption{}
\label{fig:result_basic_raw_simulator}
\end{subfigure}
\hskip2em
\begin{subfigure}{0.2\textwidth}
\includegraphics[scale=0.4]{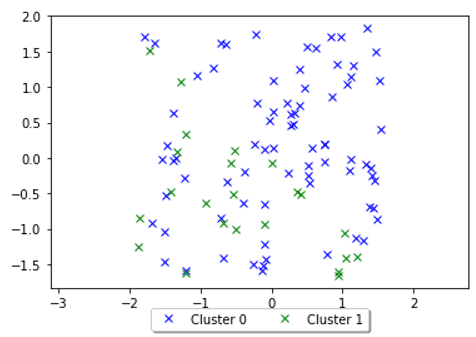}
\caption{}
\label{fig:result_basic_raw_ibmqx2}
\end{subfigure}

\begin{subfigure}{0.2\textwidth}
\includegraphics[scale=0.4]{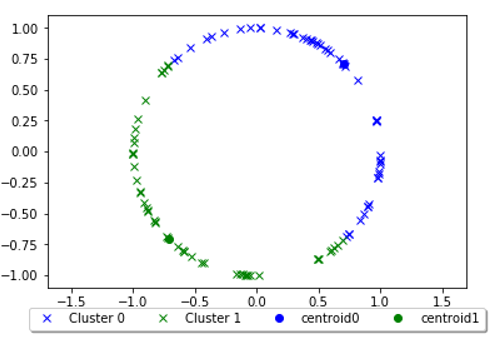}
\caption{}
\label{fig:result_basic_normalized_simulator}
\end{subfigure}
\hskip2em
\begin{subfigure}{0.2\textwidth}
\includegraphics[scale=0.4]{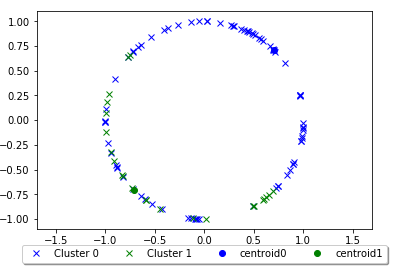}
\caption{}
\label{fig:result_basic_normalized_ibmqx2}
\end{subfigure}
\caption{Results of basic implementation on (a,c); IBM's quantum simulator, (b,d) IBMQX2.}
\label{fig:result_basic_ibm}
\end{figure}

The basic implementation can perform K-means with K = 2, which is a binary cluster finding problem. For this implementation only a random dataset (for K=2) was used to test the performance. The simulation results for this implementation are shown in Fig.~\ref{fig:result_basic_raw_simulator},~\ref{fig:result_basic_normalized_simulator}. The result shows two distinct clusters, which is the expected theoretical result for centroids $C = \{[0.707,0.707],[-0.707,-0.707]\}$. 

The results from IBMQX2 are given in Fig~\ref{fig:result_basic_raw_ibmqx2},~\ref{fig:result_basic_normalized_ibmqx2}. As can be seen from the results, the circuit performed accurately on the simulator yielding two perfect clusters that closely match the theoretical predictions, while results from the actual hardware contain considerable amount of errors. Considering simulation results as ground truth, the accuracy of the result from IBMQX2 is 54\%. The reason of these inaccurate results is the lack of fault tolerance in the current quantum hardware.  Note that most of the test vectors are assigned to cluster 0. This is because after the coherence time is over, the class qubit collapses back to its ground state $\ket{0}$ which affects the read out probabilities and it becomes difficult to get higher probability for cluster 1. 


\subsubsection{Optimized 4 qubit model}
\begin{figure}[h]
\centering

\begin{subfigure}{0.2\textwidth}
\includegraphics[scale=0.4]{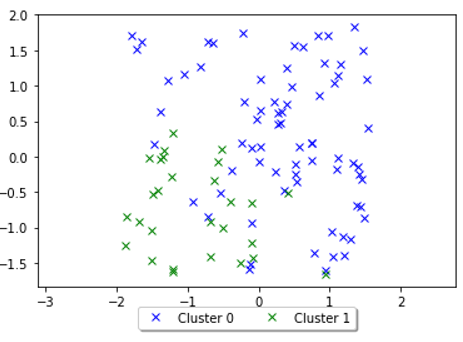}
\caption{}
\label{fig:result_optimized_raw_ibmqx2}
\end{subfigure}
\hskip2em
\begin{subfigure}{0.2\textwidth}
\includegraphics[scale=0.4]{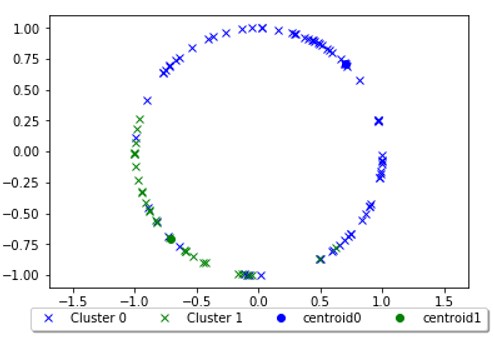}
\caption{}
\label{fig:result_optimized_normalized_ibmqx2}
\end{subfigure}
\caption{Results of optimized circuit on IBMQX2; (a): Standardized; (b): Standardized and Normalized.}
\label{fig:result_basic_ibm}
\end{figure}

The optimized circuit design for four qubits shown in Fig.~\ref{fig:optimized_interference} was first executed on IBM's quantum simulator and then on IBM's real quantum chip. Here again only the random dataset for K=2 was used. The results of the simulator are identical to the results from the original circuit. Whereas, the optimized circuit performed better in terms of accuracy on the real quantum chip. The accuracy in this case is calculated to be 76\%. These results are shown in Fig.~\ref{fig:result_optimized_raw_ibmqx2},~\ref{fig:result_optimized_normalized_ibmqx2}. Although, the results are not quite close to the simulator results, there is quite a bit of improvement when compared to the results from the circuit used in the basic implementation. The reason for this improvement is the circuit's shorter depth and reduced quantum operations.

\subsubsection{Multi-cluster Quantum K-means}
\begin{figure}[h]
\centering
\begin{subfigure}{0.2\textwidth}
\includegraphics[scale=0.4]{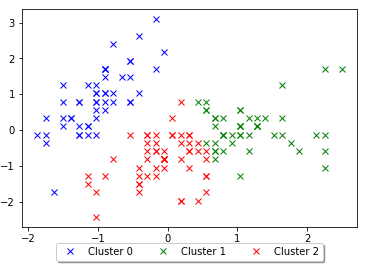}
\caption{}
\label{fig:result_scikit_iris}
\end{subfigure}
\hskip2em
\begin{subfigure}{0.2\textwidth}
\includegraphics[scale=0.4]{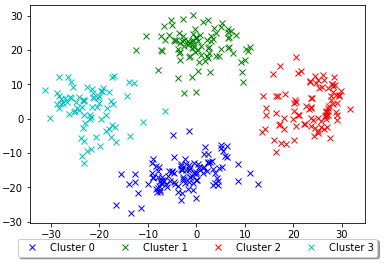}
\caption{}
\label{fig:result_mnist_scikit}
\end{subfigure}

\begin{subfigure}{0.2\textwidth}
\includegraphics[scale=0.4]{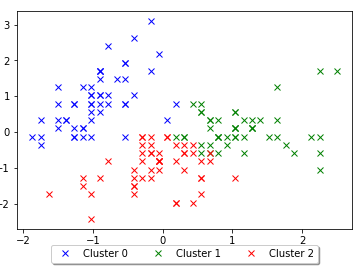}
\caption{}
\label{fig:result_iris_multi_raw_simulator}
\end{subfigure}
\hskip2em
\begin{subfigure}{0.2\textwidth}
\includegraphics[scale=0.4]{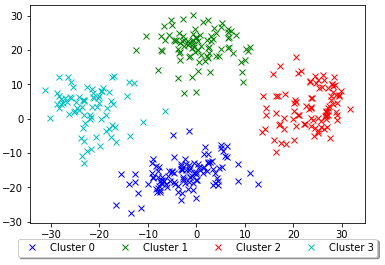}
\caption{}
\label{fig:result_mnist_simulator}
\end{subfigure}

\begin{subfigure}{0.2\textwidth}
\includegraphics[scale=0.4]{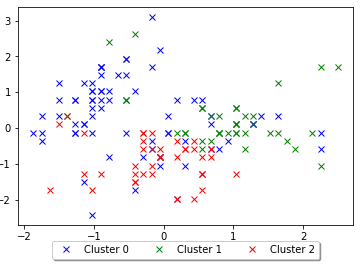}
\caption{}
\label{fig:result_iris_multi_raw_ibmqx2}
\end{subfigure}
\hskip2em
\begin{subfigure}{0.2\textwidth}
\includegraphics[scale=0.4]{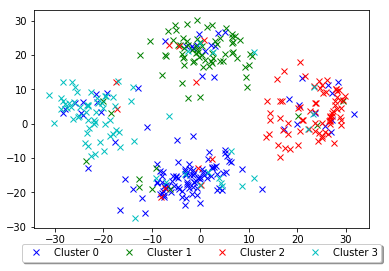}
\caption{}
\label{fig:result_mnist_ibmqx2}
\end{subfigure}

\caption{Clustering of Iris and MNIST dataset respectively using (a,b) scikit-learn kmeans on Classical computer, (c,d)  multi cluster quantum K-means on IBM's quantum simulator, (e,f) multi cluster quantum K-means on IBMQX2.}
\label{fig:result_mnist_iris_multi_sim_ibmqx2}

\end{figure}

Multi Cluster Quantum K-means strategy was tested using the Iris and MNIST datasets. Classical K-means results are also shown here for performance comparison. The accuracy is calculated using the true class labels given with the datasets. The results are shown in Fig.~\ref{fig:result_mnist_iris_multi_sim_ibmqx2}. Scikit-Learn K-means algorithm result is 83\% accurate while the accuracy for multi cluster quantum K-means is 80\%. Both algorithms are on average  99.3\% accurate when predicting the flower specie $setosa$ as it has a clear distinct cluster (see Fig.~\ref{fig:iris_dataset}). However, species $versicolor$ and $virginica$ are not linearly separable and thus can not be easily distinguished by any distance based algorithm. Therefore, when assigning clusters for test vectors from these two species both algorithms (classical and quantum) are not entirely accurate. The result of running the strategy on IBMQX2 is shown in Fig.~\ref{fig:result_iris_multi_raw_ibmqx2}. The accuracy in this case is further reduced to 62\%. This result is expected because multi cluster quantum K-means uses the same optimized quantum interference circuit which yielded 76\% accuracy on random dataset (see Fig.~\ref{fig:result_optimized_raw_ibmqx2},~\ref{fig:result_optimized_normalized_ibmqx2}). On the MNIST dataset both algorithms, scikit K-means and quantum K-means, yield identical results with 98.4\% accuracy. Note that for quantum K-means this result was produced on the simulator (see Fig.~\ref{fig:result_mnist_simulator}). When executed on the real quantum computer IBMQX2, multi cluster quantum K-means produced results with 72.6\% accuracy (see Fig.~\ref{fig:result_mnist_ibmqx2}).

\subsection{Constant Depth circuit using Negative Rotations}
\begin{figure}[h]
\centering
\begin{subfigure}{0.2\textwidth}
\includegraphics[scale=0.4]{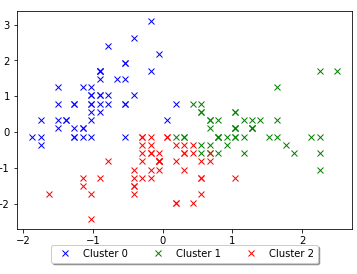}
\caption{}
\label{fig:result_iris_negative_ibmqx2}
\end{subfigure}
\hskip2em
\begin{subfigure}{0.2\textwidth}
\includegraphics[scale=0.4]{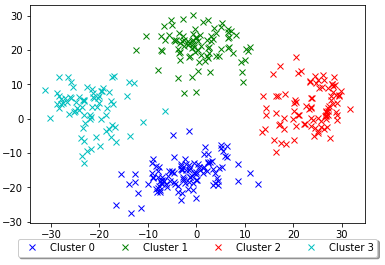}
\caption{}
\label{fig:result_mnist_negative_ibmqx2}
\end{subfigure}
\caption{Result of quantum clustering using negative rotations on IBMQX2 for (a) Iris dataset, (b) MNIST dataset.}
\label{fig:result_negative}
\end{figure}

When using the Negative rotations circuit we obtain identical solutions from both the simulation and the experimental. The depth of the circuit in this case is very low, that allows IBMQX2 to finish the computation before the coherence timeout, and hence providing the results that are theoretically predicted. The results from IBMQX2 are shown in Fig.~\ref{fig:result_negative}. 

For the Iris dataset the accuracy of the algorithm is 80\%. This is an improvement from the previous experimental accuracy on IBMQX2 calculated to be 62\% for the multi cluster strategy which uses the optimized quantum interference circuit. For the MNIST dataset the success rate is calculated to be 98.3\%. This is again an improvement from the previous result (see Fig.~\ref{fig:result_mnist_ibmqx2}) which yielded 72.6\% accuracy. 

\subsection{Distance Calculation using Destructive Interference}
\begin{figure}[h]
\centering
\begin{subfigure}{0.2\textwidth}{
\includegraphics[scale=0.4]{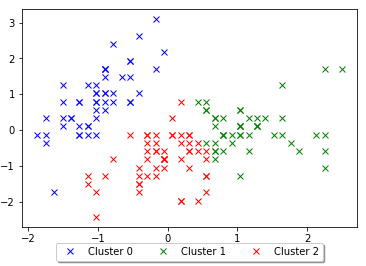}}
\caption{}
\label{fig:result_iris_destructive_ibmqx2}
\end{subfigure}
\hskip2em
\begin{subfigure}{0.2\textwidth}
\includegraphics[scale=0.4]{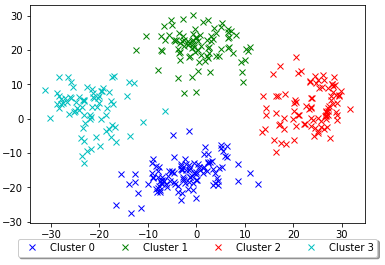}
\caption{}
\label{fig:result_mnist_destructive_ibmqx2}
\end{subfigure}
\caption{Result of quantum clustering using destructive interference on IBMQX2 for (a) Iris dataset, (b) MNIST dataset}
\label{fig:result_destructive}
\end{figure}

The circuit for distance calculation using destructive interference (see Fig.~\ref{fig:2qubit_distance}) also yielded identical results for simulated and experimental execution of the algorithm. Considering that the distance calculation is achieved with a shallow depth circuit this is to be expected. The results of running the circuit on IBMQX2 are shown in Fig.~\ref{fig:result_destructive}.

The success rate for this method is comparable to the classical scikit-learn K-means since both of these algorithms base their cluster assignments on actual Euclidean distances between vectors. For the Iris dataset the accuracy is 83.4\% and for the MNIST dataset is 98.3\%. Owing to the fact that shallow depth circuits are not swallowed up by noise on the quantum computer, these results show an improvement over the previous experimental results.

\subsection{Analysis}
Implementations with shallow depth quantum circuits are observed to be less prone to noise. The success rates for such implementations are significantly better and errors are only seen where input vectors can not be separated linearly using distance based methods. This linearly non-separable vectors are more prevalent in Iris dataset. Therefore, the maximum accuracy calculated for Iris dataset is 83.4\% which is lower when compared to accuracy on MNIST dataset using the same method.

For a complete comparison of the the different approaches discussed some additional experiments were run on IBMQX2. The results are shown in Table.~\ref{tab:Data_Combined}. The results for scikit-learn K-means are from running the algorithm locally on the classical computer.
  
\begin{table}[h]
    \centering
    \begin{tabular}{l|c|c|c|c} \hline
    & \textbf{Random} & \textbf{IRIS}  & \textbf{MNIST}  & \textbf{Circuit}  \\ 
    &  \textbf{(K=2)} & \textbf{((K=3)} &  \textbf{((K=4)} &  \textbf{(Depth} \\ \hline
    Basic &  &  &  &  \\ 
    Implementation & 54.0 & 42.0 & 31.0 & 37 \\ \hline
    Optimized   &  &  &  &  \\
     Interference  & 76.0 & 62.0 & 72.6 & 25 \\ \hline
    Negative  &   &  &  &  \\ 
     Rotations & 100  & 80.0 & 98.3 & 2 \\\hline
    Distance Calc.  &  &  &  &  \\ 
     Destructive Interference  & 100 & 83.4 & 98.3 & 14 \\ \hline
    Scikit-Learn  &  &  &  &  \\ 
     K-means & 100 & 83.0 & 98.4 & N/A \\ \hline
    \end{tabular}
    \caption{Comparison of accuracy\% for different implementations on IBMQX2. The circuit depth given is calculated by QISKIT. }
    \label{tab:Data_Combined}
    
\end{table}

The results from the basic implementation contain substantial errors. While the 4 qubit optimized interference showed considerable improvement on all datasets. Negative Rotations and Distance calculation using destructive interference methods having shallow depth circuits further improved the results.

The solutions \textit{Constant Depth Interference} \ref{sec:constant_depth_interference} and \textit{Negative Rotations} \ref{sec:negative_rotations} are limited to datasets which are based on angles. This means the datasets where magnitude of the vectors is not important for class distinction but only the angle between vectors. Polynomial feature map \cite{rebentrost2014quantum} of the quantum state can be prepared to circumvent this limitation and datasets which are not based on angles such as cocentric circles can be considered. The solution with destructive interference \ref{sec:destructive_interference} calculates actual distances between input vectors and thus it is not limited to datasets based on angles.

As a consequence of not using all quantum states $Negative\ Rotations$ only provides a linear speedup $\mathcal{O}(\frac{1}{2}NMk)$. Each qubit is used independently without using quantum entanglement. Therefore, for the generalized K-means algorithm, it scales linearly with the increasing number of $K$, vector dimensions N and number of input vectors M and requires $\frac{1}{2}NMk$ qubits.  While the other two methods discussed (\ref{sec:constant_depth_interference}, \ref{sec:destructive_interference}) use amplitude encoding, where an N dimensional vector is loaded using log N qubits, which provides the exponential speedup $\mathcal{O}(Mk(Log N))$.

\section{Conclusion} \label{sec:conclusions}
Quantum algorithms are being extensively researched seeing the potential of quantum computers to provide exponential speedups. The speedups can play a big role in machine learning where training a model is usually very slow as it requires manipulating large vectors. Quantum computers inherently are fast at manipulating and computing large vectors and tensor products. However, current quantum computers have certain limitations with respect to qubit's coherence times and noise. These barriers reduce their effectiveness on solving problems with high accuracy. In this paper, novel quantum implementations of the K-means algorithm are presented that perform clustering using shallow depth quantum circuits, which not only uses less number of quantum operations but also provide significant improvements with respect to the accuracy of the K-means algorithm.

In one solution, four qubit model (section \ref{sec:4_qubit_model}), it is discovered that the exact quantum states of the input vectors are not necessarily required for the quantum interference, if the relative angular difference between interfering vectors in the final state is equal to the interfering vectors in the input state. This finding is used to design a new quantum interference circuit with reduced number of quantum gates. In another solution which is referred to as \textit{Negative Rotations} (section \ref{sec:negative_rotations}), the cosine similarity of normalized vectors is mapped on to the probability of the qubit in state $\ket{0}$. This probability $P\ket{0}$ is used as a metric to assign clusters to input vectors. In the final solution, destructive interference (section \ref{sec:destructive_interference}) is used to calculate the actual Euclidean distance between two vectors, which is then used to perform the clustering. Experimental results show that the solutions presented provide a significant improvement in the accuracy of the K-means clustering algorithm when executed on the IBMQX2.

%

\section*{Acknowledgment}
We would like to thank Wolfgang John, Per Persson, Fetahi Wuhib and Anton Frisk Kockum for providing feedback on the first draft of the paper.

\bibliographystyle{IEEEtran}
\bibliography{IEEEabrv,refqkmeans}

\end{document}